\documentclass[twocolumn]{aastex701}
\usepackage{newtxtext,newtxmath}
\usepackage{float}
\makeatletter
\let\newfloat\newfloat@ltx
\makeatother
\usepackage{algpseudocode,algorithm}
\usepackage[utf8]{inputenc}
\usepackage{tikz}

\usetikzlibrary{fadings}
\usetikzlibrary{patterns}
\usetikzlibrary{shadows}
\usetikzlibrary{shadows.blur}
\usetikzlibrary{shapes}
\usetikzlibrary{fit}
\usetikzlibrary{calc}
\usetikzlibrary{positioning}
\usetikzlibrary{arrows}
\usetikzlibrary{arrows.meta}

\algnewcommand{\algorithmicand}{\textbf{ and }}
\algnewcommand{\algorithmicor}{\textbf{ or }}
\algnewcommand{\OR}{\algorithmicor}
\algnewcommand{\AND}{\algorithmicand}
\algnewcommand{\False}{\textbf{ false }}
\algnewcommand{\True}{\textbf{ true }}

\usepackage[T1]{fontenc}

\DeclareRobustCommand{\VAN}[3]{#2}
\let\VANthebibliography\thebibliography
\def\thebibliography{\DeclareRobustCommand{\VAN}[3]{##3}\VANthebibliography}

\usepackage{graphicx}	
\usepackage{amsmath}	

\newcommand{\hi}{H\textsc{i}}

\newcommand{\ecode}{\texttt{edges-analysis}}
\newcommand{\gscode}{\texttt{pygsdata}}
\newcommand{\gsd}{\texttt{GSData}}
\newcommand{\bowman}{B18}
\newcommand{\mahesh}{Mahesh \textit{et al.}, (\textit{in prep.})}
\newcommand{\+}[1]{\ensuremath{\boldsymbol{#1}}}

\begin{document}

\title[EDGES Analysis Pipeline]{The EDGES analysis pipeline: description and validation}


\correspondingauthor{Steven G. Murray}

\author[0000-0003-3059-3823]{Steven G. Murray}
\affiliation{Scuola Normale Superiore, Piazza dei Cavalieri 7, 56126, Pisa, Italy}
\affiliation{Physics Department, Stellenbosch University, 42 Merriman Ave, Stellenbosch, South Africa, 7600}
\email{sgmurray@sun.ac.za}

\author[0000-0003-2560-8023]{Nivedita Mahesh}
\affiliation{Cahill Center for Astronomy and Astrophysics, California Institute of Technology, Pasadena, CA 91125, USA}
\affiliation{School of Earth and Space Exploration, Arizona State University, Tempe, AZ 85287, USA}
\email{nmahesh@caltech.edu}

\author[0000-0002-6611-2668]{Akshatha K. Vydula}
\affiliation{School of Earth and Space Exploration, Arizona State University, Tempe, AZ 85287, USA}
\affiliation{Physics Department, University of Richmond, 138 UR Drive, Richmond, VA 23173, USA}
\email{akshatha.vydula@richmond.edu}

\author[0000-0002-2871-0413]{Peter Sims}
\affiliation{School of Earth and Space Exploration, Arizona State University, Tempe, AZ 85287, USA}
\affiliation{Astrophysics Group, Cavendish Laboratory, J. J. Thompson Avenue, Cambridge CB3 0HE, UK}
\affiliation{Kavli Institute for Cosmology, Madingley Road, Cambridge CB3 0HE, UK}
\email{ps550@cam.ac.uk}

\author[0000-0002-8475-2036]{Judd Bowman}
\affiliation{School of Earth and Space Exploration, Arizona State University, Tempe, AZ 85287, USA}
\email{judd.bowman@asu.edu}

\author[0000-0002-3287-2327]{Raul A. Monsalve}
\affiliation{Space Sciences Laboratory, University of California, Berkeley, CA 94720, USA}
\affiliation{School of Earth and Space Exploration, Arizona State University, Tempe, AZ 85287, USA}
\affiliation{Departamento de Ingenier\'ia El\'ectrica, Universidad Cat\'olica de la Sant\'isima Concepci\'on, Alonso de Ribera 2850, Concepci\'on, Chile}
\email{raul.monsalve@berkeley.edu}

\author[0000-0003-1941-7458]{Alan E. E. Rogers}
\affiliation{Massachusetts institute of Technology, Haystack Observatory, Westford, MA 01886, USA}
\email{arogers@mit.edu}

\author[0000-0003-0267-8432]{Rigel C. Capallo}
\affiliation{Massachusetts institute of Technology, Haystack Observatory, Westford, MA 01886, USA}
\email{rigelc@mit.edu}

\author[0000-0002-9290-0764]{John P. Barrett}
\affiliation{Massachusetts institute of Technology, Haystack Observatory, Westford, MA 01886, USA}
\email{barrettj@mit.edu}

\author[0000-0003-4062-4654]{Colin J. Lonsdale}
\affiliation{Massachusetts institute of Technology, Haystack Observatory, Westford, MA 01886, USA}
\email{cjl@mit.edu}

\begin{abstract}
The sky-averaged redshifted 21\,cm signal from Cosmic Dawn is expected to provide a unique view of the first compact objects. However, its measurement remains daunting. Difficulties are driven by the large dynamic contrast between the intervening foregrounds and the signal-of-interest, which places extremely high demands on instrumental calibration and data quality measures. The ongoing debate within the field concerning the evidence of a potential first detection by the EDGES experiment highlights the need for a more robust set of analysis methods and tools that are reliable and accessible. In this paper, we detail for the first time the precise calibration and analysis methodology adopted in previous EDGES data releases.  These methods are presented in the context of a new open-source end-to-end analysis and simulation package for 21\,cm global signal experiments that both formalizes these methods and provides general tools for the broader community.  Finally, we describe the raw data used in previous EDGES papers and release these data publicly for extended scrutiny.
\end{abstract}




\section{Introduction} \label{sec:intro}
The sky-averaged line-intensity of redshifted 21\,cm radiation from neutral hydrogen (\hi) 
is  a sought-after probe of the first billion years of cosmic history -- the Dark 
Ages ($z \gtrsim 50$), Cosmic Dawn ($50 \gtrsim z \gtrsim 12$) and Epoch of Reionization 
($12 \gtrsim z \gtrsim 5$). 
The intergalactic medium (IGM) during these epochs is dominated by neutral hydrogen, 
whose differential temperature with respect to the cosmic microwave background 
(CMB) encodes a number of key physical processes, both of cosmology and galaxy formation 
(see \cite{Furlanetto2006,Pritchard2012,Barkana2016,Mesinger2019} for excellent reviews).
To access this sky-averaged information, several low-frequency radio experiments have 
been designed and implemented over the past decade, including EDGES \citep{Bowman2008}, BIGHORNS \citep{Sokolowski2015}, LEDA \citep{Price2018a}, SARAS \citep{Girish2020,Nambissan2021}, REACH \citep{DeLeraAcedo2022}, MIST \citep{Monsalve2024}, EIGSEP \citep{Bye2025}, RHINO \citep{Bull2025}, and GINAN \citep{McKay2026}. 

Since these experiments are aimed at measuring the sky-averaged brightness temperature, 
they typically consist of a single antenna (or multiple antennas working separately) 
recording spectra over a wide frequency bandwidth (generally a 100\,MHz sub-band
in the range $\sim50-250\,{\rm MHz}$), on a cadence of several-to-tens of seconds. 
Following the strategy developed for EDGES, many of these experiments implement internal 
means of \textit{absolute calibration} to high precision. 
This cannot be achieved by externally calibrating to known sky-based sources, since the 
recorded data have little-to-no spatial resolution to isolate specific sources.

The pursuit of a measurement of the sky-averaged 21\,cm signal (hereafter 
$\bar{T}_{21}$) has been recently accelerated due to the reported observation by the EDGES collaboration 
\citep[][hereafter \bowman]{Bowman2018} of significant residual structure in the global signal after accounting for foregrounds and many instrumental systematics. This structure can be described as an absorption feature centred at 78\,MHz with a `flattened Gaussian' shape. 
\bowman\ discussed a number of hardware configuration tests that disfavor various scenarios in which this excess structure arises from non-cosmic origins.
For example, beam effects \citep{Bernardi2015} were disfavored by comparing measurements of antennas with different ground-plane sizes, and different orientations.
Conversely, polarized sky emission \citep{Spinelli2018a} was disfavored by comparing measurements from the same antenna rotated by 90$^\circ$.
Further tests also disfavour reflections of sky power from surrounding objects including the control hut, receiver calibration errors, multiplicative gains and losses and RFI.
If this excess structure is cosmic in origin, it would constitute the first direct 
evidence for the formation of the first stars at $z \sim 17$. 
However, the structure contains several
features that are difficult to reconcile with theoretical predictions. 
The depth of this feature would require either excess cooling of the IGM---beyond what is possible via 
adiabatic cooling in standard cosmological scenarios \citep{Barkana2018,Munoz2018}---
or an enhanced radio background \citep{Ewall-Wice2018,Cang2025}. 
Potentially even more difficult to reconcile with predictions is its shape; the  `sharpness' at the boundaries of the feature would require 
extremely rapid global evolution of the IGM temperature, which seems to be inconsistent 
with physical timescales of evolution on these spatial scales \citep[e.g.][]{Cang2025}. 
Furthermore, the flattened bottom of the feature is difficult to explain, as it would 
require essentially no evolution over a significant amount of time. 

However, \bowman\ and a number of following works 
\citep[e.g.][]{Hills2018a,Bradley2019,Singh2019,Sims2020} have cautioned that instrumental systematics may form a 
significant contribution to the estimated cosmic signal. 
The inherent dynamic range between radio foregrounds and the expected 21\,cm signal at 
these frequencies is very large: about $10^4$. 
While the radio foregrounds are expected to be intrinsically spectrally smooth---in 
contrast to $\bar{T}_{21}$ over a wide enough band (e.g. $50-100$ MHz)---the peculiar properties of the instrument itself introduce 
spectral structure that can obscure the signal. 
Due to the high dynamic range between foregrounds and signal, any multiplicative 
spectrally structured instrumental systematics must either be intrinsically at a level 
below $\sim 10^{-4}$, or calibrated to this same precision as part of the data analysis.
These systematics arise from a number of potential sources, including: internal reflections of the receiver \citep{Monsalve2016b,Monsalve2017,Sun2024}, 
spatial and spectral variations of the instrument's beam \citep{Tauscher2020,Sims2020,Anstey2021, Mahesh2021,Sims2023,Sims2025},
reflections of foregrounds from nearby objects \citep{Hills2018a,Rogers2022}, 
and mis-calibrated gains \citep{Murray2022b,Kirkham2025}. 
Beyond instrumental systematics, terrestrial and sky-based effects can distort incoming radiation, or add spurious power, for example Radio Frequency Interference (RFI) and ionospheric distortions \citep{Jordan2017}. 

One strategy for testing the sensitivity of the inferred signal to systematics is jackknife tests.
The large amount of EDGES data available allows for verification tests of the consistency of the cosmic signal for changes in hardware
configurations like antenna location and orientation as well as different ground planes and different processing configurations (e.g. with and without beam correction).
A number of such tests were performed in support of the results of \bowman\ (c.f. Extended Data Table 1 of that paper).

Since the number of potential systematic sources is large, and the space of 
possible perturbations of the signal due to these systematics is both large and complex, 
it is unsurprising that several techniques have been developed in an attempt to deal with them. 
Implementing these techniques requires sophisticated data-processing pipelines that make a number of 
choices: how to model calibration, identify and flag bad data, and average data together in an unbiased manner. 
Without insight into the precise choices made in these pipelines, it is incredibly 
difficult for the community to make informed judgments about the interpretations that 
proceed from the final processed data.

In this paper, we present a new data analysis package for global 21\,cm signal data, detailing the range of data analysis algorithms it provides, focusing on those used within the EDGES experiment.
This pipeline has been developed from the ground up to increase the reproducibility and transparency of EDGES' results---both its historical datasets from EDGES-2 as well as upcoming datasets from EDGES-3 \citep{Cappallo2025}.
This approach has been purposefully adopted to increase the visibility of the wider 
community into the processes underlying analysis of EDGES data, and may become more 
useful as new experiments begin to share their results.

The pipeline is extremely \textit{flexible}. Many stages of the analysis admit different approaches---for example, flagging RFI can be performed using many different techniques for identifying outlier observations---and 
here we describe the range of techniques we have implemented thus far within EDGES. 
The pipeline allows for these different algorithms and parameters to be used in a plug-and-play manner. In particular, we exemplify the pipeline by presenting the exact analysis choices made 
to produce the final averaged data product used in \bowman, and releasing all 
the raw data and pipeline code to reproduce this formative result.
We note at the outset that the data in \bowman\ were processed using an entirely different code than the one described here, which we will here refer to as the `legacy pipeline', and which is also publicly available\footnote{\url{https://github.com/edges-collab/alans-pipeline}} and minimally documented. 

The paper is organized as follows: in \S\ref{sec:overview} we give a high-level overview 
of the task of calibrating and processing sky-averaged 21\,cm data, and introduce the 
basic mathematical structure. 
In \S\ref{sec:code} we describe the software pipeline we have developed, highlighting 
aspects that are more broadly applicable to other experiments. 
In \S\ref{sec:modeling} we outline how this software handles fast linear modeling.
In \S\ref{sec:calibration} we detail how EDGES data is calibrated, in 
\S\ref{sec:flagging} how data is flagged, and in \S\ref{sec:averaging} how it is 
averaged. 
In \S\ref{sec:demo} we give an example application of the pipeline, reproducing the 
results of \bowman\ and highlighting the sensitivity of the results to various 
analysis choices. 
Finally, in \S\ref{sec:conclusions} we summarize this work.

\section{High-Level Overview} \label{sec:overview}
Sky-averaged low-frequency radio observations are measured as power spectral densities (PSDs) $P_{\rm obs}$, and we assume that they are described by
\begin{equation}
    P_{\rm obs} = g \left(T_{\rm BWsky} + T_{\rm RFI} + T_{\rm rcv} - T_{\rm loss} + N\right),    
    \label{eq:datamodel}
\end{equation}
where each component of the model is a function of both frequency $\nu$ and observation time, $t$ (but we have left these out for notational clarity).
The multiplicative gain, $g$ and the additive receiver temperature $T_{\rm rcv}$
are properties of the receiver/signal-chain that are direction-independent and applied
to the noise temperature of the antenna at the calibration plane. 
In practice, $g$ can include contributions from losses/gains inside the receiver, as well as external losses such as loss to the ground and inefficiencies in the antenna. 
In general, losses have both a multiplicative and additive component, and $T_{\rm loss}$ represents the additive component.
Conversely, the beam-weighted sky, $T_{\rm BWsky}$, and $T_{\rm RFI}$ are intrinsically 
direction-dependent properties of the sky that are integrated over the horizon-to-horizon
field-of-view of the instrument, with a weighting defined by the primary beam of the
antenna:
\begin{equation}
    T_{\rm BWsky} + T_{\rm RFI} = \int_{2\pi} d\Omega\ B(\nu, \Omega) \left[ T_{\rm sky}(\nu, \Omega, t) + T_{\rm RFI}(\nu, \Omega, t)\right].
\end{equation}
The final term in Eq. \ref{eq:datamodel}, $N$, is the thermal noise.

In this work we will consider the beam-weighted sky temperature to be the sum of the global 21-cm temperature and another term representing all (beam-weighted) foregrounds:
\begin{equation}
    T_{\rm BWsky} = g_{\rm beam}(\nu, t) \left[ T_{21}(\nu) + T_{\rm BWFG}(\nu,t)\right],
\end{equation}
where $g_{\rm beam}$ represents the `chromaticity' of the beam (i.e. the extra factor that arises due to $B$ being dependent on $\nu$, rather than simply $\Omega$).
We write $T_{21}$ instead of `$T_{\rm BW21}$' because the 21\,cm signal is expected to be homogeneous on the scale of the beam size. 
Furthermore, due to this homogeneity, $T_{21}$ is not time-dependent.

Since real observations are discrete in time and frequency, we write the various temperatures and observed power instead as data vectors with length given by the product of number of frequency channels by number of observed integration times, $N_\nu N_{\rm intg}$:
\begin{equation}
    \+P_{\rm obs} = \mathbf{G} \left(\mathbf{G}_{\rm beam} \left[\+T_{21} + \+T_{\rm BWFG}\right] + \+T_{\rm RFI} + \+T_{\rm rcv} - \+T_{\rm loss} + \+N\right),    
    \label{eq:datamodel-linear}
\end{equation}
Here, italic-bold represents a vector, upright-bold $\mathbf{G}$ indicates a matrix, and in practice $\mathbf{G}$ is diagonal.

We assume the noise $\+N$ to be Gaussian-distributed with a mean of zero and a diagonal covariance whose diagonal terms are given by the radiometer equation:
\begin{equation}
    \sigma^2_T \approx \frac{(T_{\rm BWsky} + T_{\rm rcv} + T_{\rm RFI} + T_{\rm loss})^2}{\Delta \nu \Delta t}.
\end{equation}
In detail, the spectrometer correlates neighbouring frequency channels and times. However, this correlation is negligible for practical applications.

Our formulation in Eq.~\ref{eq:datamodel-linear} supports our goal in the 
EDGES analysis pipeline, which is to determine an unbiased estimate of the true spectrum of the sky, $\hat{\+T}_{\rm BWsky}$ (ultimately, the goal is to produce unbiased estimates of $\+T_{21}$, but the careful separation of the foregrounds from the 21-cm signal is more properly within the scope of \textit{inference} rather than \textit{analysis} and is thus beyond the scope of this paper)\footnote{Technically, all analysis steps presented in this paper should also properly be considered as part of inference, which is a long-term goal of this project, however the strong correlations between foreground and 21-cm models makes it all the more necessary when separating them}.
There are three aspects of this process:
\begin{enumerate}
    \item \textit{Calibration} refers to the estimation of $\mathbf{G}$, $\mathbf{G}_{\rm beam}$, $\+T_{\rm rcv}$ and $\+T_{\rm loss}$ such that the observed PSD can be calibrated to a temperature.
    \item \textit{Flagging} refers to identification of either specific channels or 
          integrations that are afflicted by sources not described by any other component
          of the model, which we have here grouped into the catch-all term 
          $T_{\rm RFI}$\footnote{Note that terrestrial radio-frequency interference (RFI)
           is not the only source of aberrant power that causes the observed data to 
           deviate from our model, however since it is the most common we let it stand
           in for all sources of such extra power.}.
          Instead of attempting to determine the magnitude of the undesirable sources 
          of power, we opt to simply identify its presence and excise such channels and
          integrations from the data entirely. 
    \item \textit{Averaging} refers to combining either channels or integrations with the aim of reducing the variance of $N$ without significnatly reducing information in (or biasing) the global 21-cm signal, $\+T_{21}$. 
\end{enumerate}

Given a dataset $\+P_{\rm obs}$ (which is a vector of length $N^{\rm raw}_{\rm \nu}N^{\rm raw}_{\rm intg}$), the analysis process can be written as a series of linear operations, each of which corresponds to one of these these three aspects (each can be applied multiple times). Letting the intermediate result at any point $k$ in this process be $\+X_k$, with $\+X_0 = \+P_{\rm obs}$, each step of the process can be written\footnote{
Note that while this form (Eq. \ref{eq:high-level-data-reduce}) seems linear in the data $\+P_{\rm obs}$, it is often not linear in detail because the dataset itself can be used to determine $\mathbf{\mathbf{\Xi}}$.}:
\begin{equation}
    \+X_{k+1} = \mathbf{D}_k \mathbf{S}_k (1 - \mathbf{\Xi}_k) \hat{\mathbf{G}}^{-1}_k \+X_k - \hat{\+T}_{{\rm cal}, k}.
    \label{eq:high-level-data-reduce}
\end{equation}
In this equation, the $\mathbf{G}_k$ are square, diagonal matrices representing estimated \textit{calibration} gains (potentially including receiver gains, cable losses, beam chromaticity corrections and any other multiplicative correction), and $\hat{\+T}_{{\rm cal},k}$ are estimated calibration temperatures (i.e. any additive correction).
In practice, for most analysis steps no calibration will be applied, in which case $\mathbf{G}_k = \mathbf{I}$ and $\hat{\+T}_k = 0$.
\textit{Flagging} is applied through $\mathbf{\Xi}_k$, which are square diagonal matrices of zeros (good data) and ones (flags).
\textit{Averaging} is achieved through the application of $\mathbf{D}_k\mathbf{S}_k$, where $\mathbf{S}_k$ are square smoothing kernel matrices whose rows sum to unity, and $\mathbf{D}_k$ are rectangular down-sampling matrices, in which each row contains a single non-zero entry of unity (in a step in which no averaging is performed, their product is simply set to the identity matrix).

Eq. \ref{eq:high-level-data-reduce} represents an extremely high-level formulation of the general data processing pipeline, resulting in $\+X_n = \hat{\+T}_{\rm BWsky}$. 
The aim of this paper is to present methods for determining $\mathbf{D}_k$, $\mathbf{S}_k$, $\mathbf{\Xi}_k$, $\mathbf{G}_k$, $\+T_{\rm loss}$ and $\+T_{\rm rcv}$ based on characteristics of the EDGES data, but also broadly applicable to 21\,cm global-signal experiments.

\section{Software for Data Interfaces and Analysis} \label{sec:code}
Given the extreme requirements on calibration precision for an unambiguous detection of the global 21\,cm signal, it is paramount that any uncertainties stemming from analysis choices are well-characterized.
Analysis pipelines required for global signal data are full of impact-laden choices---especially as regards to how and at what stage the data are flagged for particular effects, and the process of modeling (or `inpainting') to achieve unbiased averaging. 
It is therefore crucial that the analysis pipeline software lends itself to easy exploration of different parameter and algorithm choices, at the same time as ensuring that these choices are traced for reproducibility.
This design approach is made all the more necessary by the looming threat of methodological confirmation bias---i.e. the tendency to choose parameters/methods that yield the result that is either expected or desired, rather than purely due to their intrinsic merit---, especially given the high-profile results from EDGES-2.
Furthermore, it is important that the analysis software is well-documented and transparent, so that the community can validate the procedure. 

The software used to generate the results presented in B18 was neither modular, open-source, nor easily traceable. It was originally developed for internal use rather than for broader accessibility or reproducibility. Here, we present a new suite of Python packages, developed under the \texttt{edges-collab} Github organization\footnote{\url{https://github.com/edges-collab/}} that meet the challenges of modularity, reproducibility, and accessibility we have just described. The packages of most interest to the broader community are \texttt{pygsdata}\footnote{\url{https://github.com/edges-collab/pygsdata}}, which provides a consistent unified interface to low-frequency single-antenna radio data, and \ecode, which provides a number of analysis tools built on that core data-interface. The high-level goals of these packages are to be broadly applicable (i.e., able to be used for experiments outside EDGES), modular (i.e., easy to switch between different methodological choices in a plug-and-play fashion), accessible (open-source and extensively documented using modern software design principles to enable the code to be easily understood), reliable (unit-tested with over 90\% code coverage as well as end-to-end tested against the legacy software), and performant. 

The reason \texttt{pygsdata} stands separate from \ecode\ is that it is entirely experiment-agnostic. 
The \texttt{pygsdata} package defines a single data interface that captures the range of data shapes across global experiments (e.g., multiple antenna inputs and polarizations) and unifies common tasks like data selection, concatenation and reading/writing. This interface is intended to ease the sharing of data between experiments, and potentially also the sharing of methodologies. 
It adopts a similar motivation and design philosophy as the \texttt{pyuvdata}\footnote{\url{https://github.com/RadioAstronomySoftwareGroup/pyuvdata}} project does for interferometric data \citep{Hazelton2017,Keating2025}.
While \ecode\ is by necessity slightly more geared towards EDGES data (for example, focusing on calibration using the noise-wave formalism of \citealt{Rogers2012} and \citealt{Monsalve2017}), its intrinsic modularity means that many components are much more widely applicable and may easily be incorporated into the analysis pipelines of other experiments.
In particular, it fundamentally rests on the core \texttt{pygsdata} interface, which means that many of the functions will be easily applied to data from other experiments. 

\subsection{The Data Interface}
\label{sec:code:pygsdata}
The \texttt{pygsdata} package is a best-principles software package that defines Python data-types for interfacing with low-frequency radio single-antenna spectral data.
The fundamental data-type in \texttt{pygsdata} is a \texttt{GSData} object.
At its core, the spectrum data is encoded with the \texttt{GSData} object as a four-dimensional array of shape $(N_{\rm loads}, N_{\rm pols}, N_{\rm times}, N_{\rm freqs})$.
Along with the spectrum data, two other arrays of the same shape are permitted, specifically an array holding the number of samples comprising each datum, and an array of residuals of the data to a model. 
Besides these data arrays, the \texttt{GSData} object holds several important pieces of metadata. These include the metadata associated with the axes of the data: an array of times, an array of frequencies, the list of polarizations and the list of load-names. 
Other metadata include the telescope name and location, as well as a table of `auxiliary measurements' which can be arbitrary quantities measured at each observed time (these can be used to track important corollary measurements such as ambient temperature). 

Unlike \texttt{pyuvdata}, which formed a large inspiration for \gscode, \gsd\ objects are \textit{immutable}: they are initialized and validated at a single point and then remain forever valid afterwards.
This is a significant advantage for analysts using the data, as it makes keeping track of the state of the object much more simple. 

Instead of updating parts of the data object in-place via a plethora of attached methods, the \gsd\ interface is extremely simple, having only a few essential methods, generally provided for easily accessing slices of the data (or other data-like attributes), and selecting out parts of the data (frequency range, Local Sidereal Time (LST) range, etc) into a new valid \gsd\ object. 
Instead, new transformed objects (down-selected, combined, calibrated, flagged and averaged, c.f. \S\ref{sec:overview}) are created from input objects  via simple external functions.
As part of this architecture, \gscode\ provides a decorator-based mechanism for registering new functions whose primary purpose is to ingest \gsd\ objects and return new modified ones. 
While any function can do this, registering a \gsd-analysis function provides benefits such as automatically tracking the history of the data.

The \gsd\ object contains a \texttt{.history} attribute which is essentially a list of analysis-stamps, each of which contains metadata about a function that has been applied to the data (the time of application, the name of the function, versions of key packages involved, and the parameters passed to the function). The \gsd\ object has methods for inspecting and printing this history.
This means that the final processed spectrum from a given analysis has a full traceable history of all the processes applied to it, essential for reproducibility. 

\gscode\ makes full use of modern packages such as \texttt{astropy} \citep{AstropyCollaboration2013,AstropyCollaboration2018}.
For instance, all applicable attributes of the \gsd\ object are Quantities with units, which reduces ambiguity and enables dimensional analysis.
Furthermore, functionality from \texttt{astropy} is used to determine the LSTs corresponding to observation times, and to identify the sky that the telescope can view at any particular time. 

\gsd\ also has an advanced system for storing and applying flags. 
It is common to store flags as a single boolean array of the same shape as the data. 
However, this means that flags arising from different quality checks cannot be distinguished after they are combined into the single array, forfeiting the ability to later compare the impact of multiple quality checks. 
In \gscode\, flags from a single quality check are stored in a bespoke format---a \texttt{GSFlag} object---which is a boolean array with up to four dimensions, where the quantity corresponding to each axis (e.g. polarization, time, frequency) is explicitly specified.
It is common in the analysis of EDGES data to flag entire spectra at once (i.e. every channel for a particular time).
The format of \texttt{GSFlag} means that this can be stored as a small array of size $N_{\rm times}$ instead of a full 4D array.
Within the \gsd\ object, the flags are stored as a mapping of quality-check names to \texttt{GSFlag} objects.
The full resulting set of flags, combined via boolean `OR' (consistently combining flags with different dimensionalities), is made available as the \texttt{.complete\_flags} attribute. 

\gsd\ is intended to be a unified interface for all single-antenna radio data. To that end, \gscode\ defines a HDF5-based self-describing data format for storing the data on disk. 
While this format is feature-complete and fast both to read as a \gsd\ object and to write from such an object, almost all experiments have developed bespoke formats for their spectrometer data. 
Indeed, even EDGES natively writes data in a bespoke  format using \textit{uuencoding} (binary-to-text encoding) internally referred to as `ACQ' files.
To enable various groups to easily adopt \gsd\ for their data, \gscode\ provides a decorator-based registration mechanism so that arbitrary functions can be defined that read a particular data format and return a \gsd\ object. When registering a custom reader, a set of file-suffices can be registered as associated with the reader, so that simply importing the reader function makes it possible to use the built-in constructor \texttt{GSData.from\_file} to read any file with that suffix. 
The \texttt{read-acq} package uses just this mechanism to enable reading of \texttt{.acq} files produced by the EDGES spectrometer.

\subsection{The Analysis Package}
\label{sec:code:analysis}
The \ecode\ package comprises a number of sub-packages that focus on different aspects of data processing.
Some of the more important sub-packages of \ecode\,, most of which we will describe in more detail in the following sections, include 
\begin{itemize}
    \item \texttt{edges.modeling} (c.f. \S\ref{sec:modeling}), an interface for defining and fitting linear models;
    \item \texttt{edges.io}, routines for reading and writing data formats other than the spectra themselves;
    \item \texttt{edges.filters} (c.f. \S\ref{sec:flagging}), a set of quality checks and data filters (e.g. RFI) that apply to \gsd\ objects;
    \item \texttt{edges.averaging} (c.f. \S\ref{sec:averaging}), methods for unbiased averaging and/or binning over different data axes;
    \item \texttt{edges.cal} (c.f. \S\ref{sec:calibration}), methods to calibrate and model reflection coefficients and determine receiver calibration solutions;
    \item \texttt{edges.analysis}, high-level methods for applying the algorithms of other sub-packages to \gsd\ objects; and
    \item \texttt{edges.inference}, containing non-linear models of both foregrounds and 21\,cm signals, and a unified interface for constraining their parameters via Bayesian inference.
\end{itemize}

While some of these sub-packages are more-or-less tailored for EDGES data (e.g \texttt{edges.io}), most provide standalone methods that are broadly applicable, as they are based on the unified \gsd\ interface. 
For example, the \texttt{edges.filters} sub-package provides a number of filters for data quality (c.f. \S\ref{sec:flagging}), all of which accept a \gsd\ object and return a new one, duly flagged. 
Examples of general-purpose filters are those that flag data dependent on the altitude of the sun/moon, those that flag integrations with abnormally high power, and a number of RFI filters that search for compact outliers using different algorithms. With this software, the construction of full end-to-end analysis pipelines is generally quite simple, predominantly consisting of a series of function calls that sequentially evolve a \gsd\ object. 
The resulting pipeline is thus also highly transparent. 

\section{Linear Modeling} \label{sec:modeling}
Many aspects of the analysis of single-antenna radio data call for modeling the data as smooth functions (typically of frequency). 
For instance, such models can be used to reduce noise in VNA measurements while simultaneously interpolating the measurements to the channels at which the spectral data were taken.
These models can also be used to in-paint flagged data in order to mitigate biasing when averaging (c.f. \S\ref{sec:averaging}), help to identify outliers for flagging (c.f. \S\ref{sec:flagging:rfi}) and form physical models of foregrounds when interpreting the data.

The \texttt{edges.modeling} sub-package provides a convenient, flexible, composable and fast interface for defining, evaluating and fitting linear models to data.
While linear modeling tools are widely available (e.g. in recent versions of \texttt{astropy}\footnote{https://docs.astropy.org/en/latest/modeling/index.html}, as well as lower-level tools such as the linear-algebra routines and polynomial fitting in \texttt{numpy}), the \texttt{edges.modeling} package was written to enable the following features:
\begin{itemize}
    \item The ability to define models with arbitrary numbers of terms (e.g. polynomial or Fourier basis sets with $n$ terms).
    \item Easy serialization/definition of the models to human-readable formats like YAML.
    \item Ability to cache the basis functions to make repeated fits faster.
    \item Ability to apply transformations to the coordinates at which the model terms are evaluated, separate from the definition of the model (e.g. normalising frequencies by some reference frequency, $x=\nu/\nu_{\rm ref}$ and evaluating polynomials in terms of $x$).
    \item Unified support for multiple fitting methods, including support for varying weights.
\end{itemize}

In \texttt{edges.modeling}, a model is defined by specifying a functional form for each term:
\begin{equation}
    M = \{m_0(x), m_1(x), \dots, m_{n-1}(x) \},
\end{equation}
where $x = \mathcal{T}(x')$ is a vector of coordinates at which the term is evaluated (transformed from the natural coordinates $x'$ via the user-specified transformation $\mathcal{T}$). 
Many models are defined for an arbitrary number of terms, and the specific number of terms required for any particular application is set by the analyst when using the model.
Other models are only defined for a certain number of terms, or a certain range of terms. 

Models are evaluated for a particular set of coefficients, $\+a = \{a_0, a_1, \dots, a_{n-1}\}$, at a particular vector of coordinates $\+x = \mathcal{T}(\+x')$, by packing the vectors $\+m_i$ into an $N_x \times n$ matrix $\mathbf{M}$ and computing $\mathbf{M}\+a$.
Conversely, the parameters are fit to the data vector $\+y$ using linear least-squares (potentially with non-uniform weights) via one of several matrix inversion methods (e.g. QR-decomposition).

\subsection{Common Models}
\label{sec:modeling:models}
There are a number of linear models that are commonly used in analysis of global signal data, and these are built-in to \texttt{edges.modeling} for convenience. We do not present all of them here, but specifically note those that will be used throughout this work.

Common useful flexible models include the Fourier model:
\begin{equation}
    M_{{\rm fourier}, k} = \begin{cases}
        \sin\left(2\pi k x\right) & k \geq 0 \\
        \cos\left(2\pi k x\right) & k < 0 ,
    \end{cases}
    \label{eq:models:fourier}
\end{equation}
for $(1-n)/2 \leq k \leq (n-1)/2$.
Another flexible model is the modified polynomial:
\begin{equation}
    M_{{\rm poly}, k} = x^{pk - q},
    \label{eq:models:poly}
\end{equation}
where $q$ is a parameter that can be set to more easily capture behaviour close to a negative power-law (common for foreground spectra as a function of frequency), and $p$ is a parameter controlling how much extra structure is introduced in each new term.

We also include several models that are designed to capture foreground structure with a low number of terms. In these models, $\beta$ is a parameter of the model that generally sets the dominant spectral index of the foreground spectrum. These include a linearized form of a physically-motivated foreground model including ionospheric terms, \textsc{LinPhys}:
\begin{equation}
    M_{\rm linphys} = \{x^\beta, x^\beta \log x, x^\beta \log^2x, x^{\beta-2}, x^{-2} \},
    \label{eq:models:linphys}
\end{equation}
a polynomial in log-frequency, \textsc{LinLog}:
\begin{equation}
    M_{{\rm linlog}, k} = x^\beta \log^k(x),
    \label{eq:models:linlog}
\end{equation}
and the so-called \textsc{EdgesPoly}:
\begin{equation}
    M_{{\rm epoly}, k} = M_{{\rm poly}, k}(p=1, q=\beta).
    \label{eq:models:epoly}
\end{equation}

\section{Calibration} \label{sec:calibration}
The calibration method for EDGES data was presented in detail in \citep{Monsalve2017} and has been elaborated in \citep{Murray2022b} in the context of a Bayesian formalism.
The method is a two-step process, in which time variability is removed with a Dicke-switching procedure, while absolute temperature calibration is performed using thermal references and the noise-wave formalism \citep{Meys1978}.
Given the success of this approach, it has been adopted by a number of other experiments, including REACH \citep{Roque2021,DeLeraAcedo2022}, MIST \citep{Monsalve2024}, LEDA \citep{Price2018a} and RHINO \citep[][though they also use a signal-tone injection calibration scheme on mid-range time-scales]{Bull2025}.
Thus, the brief outline of the calibration formalism presented here is broadly applicable.

Both EDGES-2 and EDGES-3 follow the formalism outlined in this section, though some of the details are different between the implementations. 
For instance, EDGES-3 uses physically modified versions of the same kinds of calibration sources as EDGES-2, but they are connected through a switch network internal to the receiver box, rather than external sources used in a lab setting \citep{Cappallo2025}. 
For brevity (and to support the demonstration in \S\ref{sec:demo}), in this work we will refer only to the calibration of EDGES-2, though \ecode\ is modular enough to support calibration of both systems without much effort.

There is much publicly accessible documentation describing details of this procedure, especially in the series of memos hosted at both ASU\footnote{\url{https://loco.lab.asu.edu/memos/}} and MIT\footnote{\url{https://www.haystack.mit.edu/haystack-memo-series/edges-memos/}}.
Our purpose in this paper is to provide a comprehensive overview regarding the mathematical framework and algorithms involved in calibration, rather than a thorough description of the procedural aspects of taking the measurements. 
For the latter, see e.g. \citet{Mozdzen2018}.
At the same time, we aim here to frame some of the more technical aspects that involve electrical-engineering concepts in a more conceptual way for analysis practitioners with less familiarity in this area.

\subsection{The EDGES-2 System}
\label{sec:calibration:system}

\begin{figure}
    \centering
    \includegraphics[width=\linewidth,trim=18cm 3cm 18cm 3cm,clip]{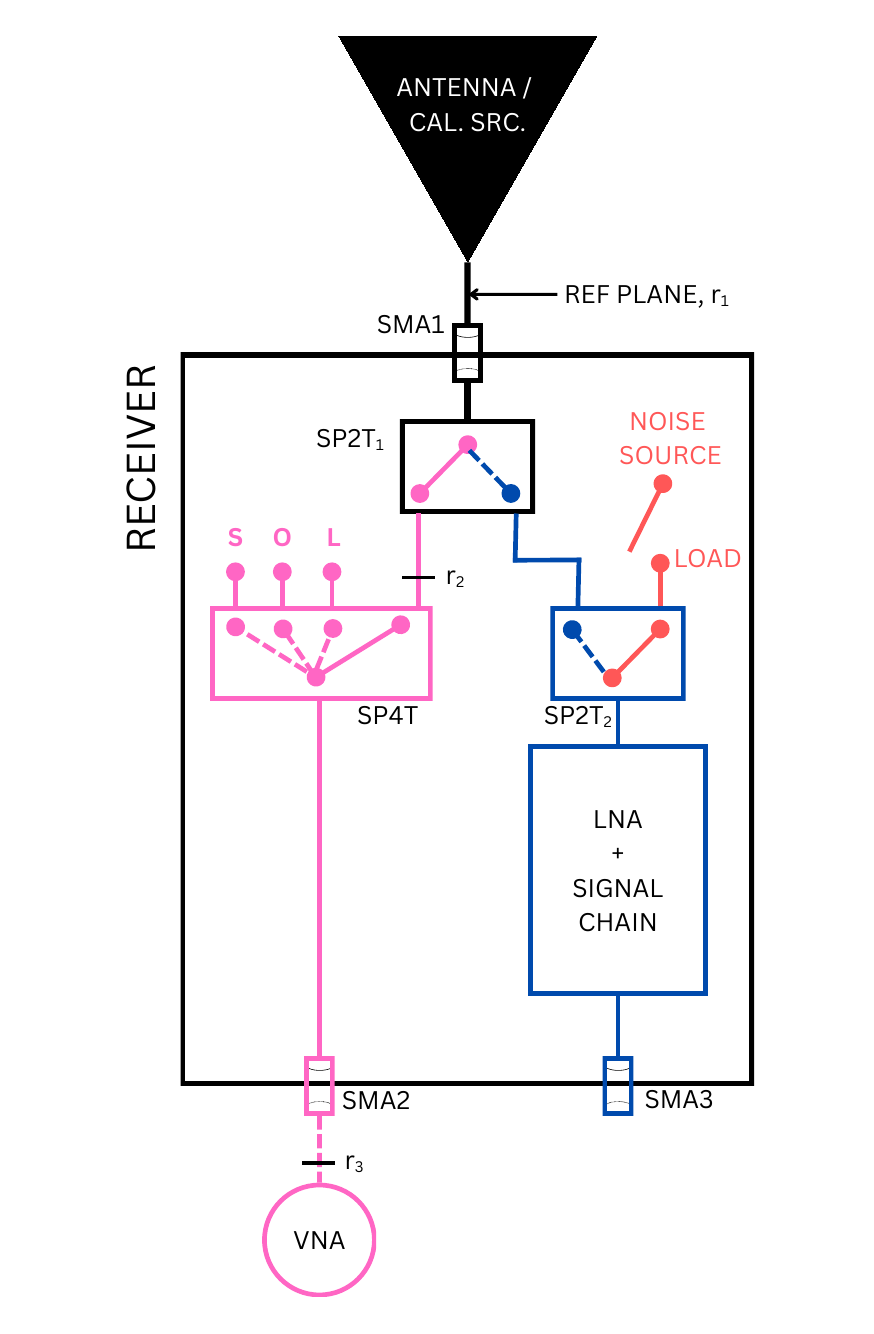}
    \caption{System diagram for the EDGES-2 receiver. We refer to the entire enclosed box as the `receiver'. The pink components define the internal subsystem for measuring and calibrating input source reflection coefficients. The blue components define the main signal path for recording spectra from the antenna/calibration source. The orange components comprise the internal system for relative calibration using Dicke switching (c.f. \S\ref{sec:calibration:dicke}). }
    \label{fig:edges2-diagram}
\end{figure}

Fig. \ref{fig:edges2-diagram} gives a high-level schematic overview of the EDGES-2 system. 
In this diagram, several of the annotated components are multi-component subsystems. 
For the purposes of this description of the calibration algorithms, these subsystems can be regarded as black boxes.

The system begins with an input source, which is generally the antenna itself or some other electrical device like a passive calibration source.
This source is connected to the receiver box via an SMA connector\footnote{Subminiature version A (SMA) connectors are standard coaxial connectors for radio-frequency signals, with a nominal impedance of 50\,$\Omega$.} (SMA1), and from here signals can follow two main routes: either through a four-position switch to a vector network analyzer (VNA) connected via another SMA (SMA2) for injecting signals to measure internal reflections (pink route); 
or through the low-noise amplifier (LNA) and rest of the signal chain to an output SMA (SMA3) for connection to the spectrometer (blue route). 
In the field, the main two-position switch (SP2T$_1$) is generally connected to the blue route in order to record spectra, but is occasionally switched over to the pink VNA route for updated measurements of the antenna reflection. A final subsystem is an internal noise-generating resistive load implemented as an attenuator, as well as an active noise source (orange route) that boosts the noise exiting the load.
The system intermittently switches between these internal calibration loads and the input source via another two-position switch (SP2T$_2$, c.f. \S\ref{sec:calibration:dicke}).                                                        
An important point is that the \textit{reference plane}---the location on the diagram at which reflections are referenced---is at the input of SMA1 (or equivalently, the output of the input source). 
That is, all relevant reflection coefficients are defined as the ratio of the current returned to the reference plane from the connected source (e.g. antenna) in response to an injected current at this reference plane from the VNA (c.f. Fig. \ref{fig:vna-schematic}). 
Much of the procedure for calibrating the system is concerned with translating measurements of reflections at other points in the system back to this reference plane.

\subsection{Dicke Switching}
\label{sec:calibration:dicke}
Dicke-switching works by switching the internal input from the antenna feed to two different calibration sources sequentially on a cadence of approximately 13 seconds each\footnote{As pointed out in \citet{Bull2025}, it can be advantageous to have the integration time for each internal Dicke-switch source be different, but this does not affect the formalism as presented here.}.
In the EDGES systems, these sources are a `load' which is at ambient temperature, and a `load plus noise-source' for which the noise-source contributes an additional signal equivalent to a noise-source well above ambient temperature (see Fig. \ref{fig:edges2-diagram}).
For each input $X \in \{{\rm ant}, {\rm L}, {\rm L+NS}\}$ we measure the PSD, i.e. $P_X = g_X (T_X + T_{X, {\rm off}})$ (c.f. Eq. \ref{eq:datamodel}).
For each 39-second cycle, these three inputs are combined to form the following quotient:
\begin{equation}
    Q  = \frac{P_{\rm ant} - P_{\rm L}}{P_{\rm L+NS} - P_{\rm L}}.
    \label{eq:dickecal}
\end{equation}
To first order, since the PSD measured at each switch position experiences the same gain $g_X \approx g_{\rm rcv}$ and additive temperature $T_{X, {\rm off}} \approx T_{\rm rcv}$ from the receiver subsystem, this quotient is approximately\footnote{Note that there is a small bias that we neglect in this equation which arises due to the statistical correlation of the numerator and denominator in Eq. \ref{eq:dickecal}---since both contain the same $P_L$ observations. This however is very small ($\sim10^{-5}$\,K) for the long integration times we adopt on each cycle \citep[e.g.][]{Rogers2015}.} 
\begin{equation}
    Q \approx \frac{T_{\rm ant} - T_{\rm L}}{T_{\rm NS}}.
    \label{eq:q-basic}
\end{equation}
The physical temperature of the receiver is controlled and stabilized to high accuracy.
This stabilization enables us to assume that $T_{\rm L}$ and $T_{\rm NS}$ are time-independent.
Thus, all we need is to measure $T_{\rm L}$ and $T_{\rm NS}$ once, and we can invert Eq. \ref{eq:q-basic} to obtain $T_{\rm ant}$.
In practice, there are a few complications to this picture; 
these complications centre around the fact that the signal path from the antenna to the receiver is different than that between the internal loads and the receiver, as it includes an extra two-position switch (SP2T$_1$).
Furthermore, the impedance mismatch between the antenna and the receiver causes internal reflections that need to be taken into account.
This impedance mismatch requires incorporating the noise-wave formalism, which we will describe next.
Nevertheless, even with the added complexity of the noise-waves, the ultimate calibration from $Q$ to $T_{\rm ant}$ remains linear and can be expressed through two calibration temperatures: a scaling and offset temperature\footnote{Note that the models for these scaling and offset temperatures may not be linear in the parameters chosen to represent them, especially if the internal load is not well matched, but the relation between $T_{\rm obs}$ and $Q$ remains linear.}:
\begin{align}
    \label{eq:TscaToff}
    T_{\rm ant} &= T_{\rm sca} Q + T_{\rm off}  \nonumber \\  
    T_{\rm sca} &\approx T_{\rm NS} \\
    T_{\rm off} &\approx T_{\rm L}. \nonumber
\end{align}

\subsection{The Noise-Wave Formalism}
\label{eq:noise-waves}
For any source connected to the LNA, the output power from the receiver can be written as
\begin{align}
    \label{eq:psrc_full}
    P_{\rm src} = g_{\rm src} \big[ & (1 - |\Gamma_{\rm src}|^2) F_{\rm src}^2 (1 - L_{\rm src}) T_{\rm src}  \nonumber\\
    & + L_{\rm src} T_{\rm amb} \nonumber\\
    & + |\Gamma_{\rm src}|^2 |F_{\rm src}|^2 T_{\rm unc} \nonumber \\
    & + |\Gamma_{\rm src}| |F_{\rm src}| \cos \alpha_{\rm src} T_{\rm cos} \nonumber \\
    & + |\Gamma_{\rm src}| |F_{\rm src}| \sin \alpha_{\rm src} T_{\rm sin} \nonumber \\
    & + T_{\rm rcv}\big] .
\end{align}
Here $\Gamma_{\rm src}$ is the complex-valued reflection coefficient (i.e. $S_{11}$) of the source and $\Gamma_{\rm rcv}$ is its counterpart for the receiver input. 
$F$ and $\alpha$ are functions of the reflections coefficients:
\begin{align}
    F_{\rm src} = \frac{\sqrt{1 - |\Gamma_{\rm rcv}|^2}}{1 - \Gamma_{\rm rcv}\Gamma_{\rm src}}, \\
    \alpha_{\rm src} = {\rm arg}(\Gamma_{\rm src} F_{\rm src}).
\end{align}
In Eq. \ref{eq:psrc_full}, $L_{\rm src}$ quantifies the \textit{loss} experienced by the input temperature before it reaches the SMA1 input connector (e.g. if the source device is connected via a cable).
Its value is zero for lossless systems and one for those with total loss. 
The second term, $L_{\rm src}T_{\rm amb}$, is an additive temperature arising from the ambient temperature of the lossy input device (c.f. \S\ref{sec:calibration:source-loss}).

In Eq. \ref{eq:psrc_full}, the receiver gain $g$ is independent of the input source for a fixed calibration plane. 
However, since the path from the input source to the LNA is different from the internal loads to the LNA, $g$ is in general different between the two (but not between the two internal loads).                                                                                                                      
For the internal loads, the reflection coefficient $|\Gamma_{\rm src}|$ is negligibly small since they are very close to 50$\Omega$ by construction.
We can thus approximate $|\Gamma_{\rm L}| = |\Gamma_{\rm LNS}| \approx 0$ (see \citealt{Kirkham2025}, their Section 2.2, for a discussion of impedance mismatch in the internal references). Furthermore, as modeled, in the internal loads there is negligible loss, so that the internal load powers may be written:
\begin{align}
    P_{\rm L} &= g(1 + \delta_g) (1 - |\Gamma_{\rm src}|^2) \left[T_{\rm L} + T_{\rm rcv}\right], \\
    P_{\rm L+NS} &= g(1 + \delta_g) (1 - |\Gamma_{\rm src}|^2) \left[T_{\rm L} + T_{\rm NS} + T_{\rm rcv}\right].
\end{align}

Writing $T'_{\rm L} = (1 + \delta_g)T_{\rm L}$ and $T'_{\rm NS} = (1 + \delta_g)T_{\rm NS}$, 
we can then rewrite the quotient for some input source as 
\begin{equation}
    Q_{\rm src} = \frac{\big[\epsilon_{\rm src} T_{\rm src}
     + L_{\rm src}T_{\rm amb}
     + \+{K} \cdot \+{T}_{\rm NW}
      - (1 - |\Gamma_{\rm rcv}|^2) T'_{\rm L} \big]}{(1-|\Gamma_{\rm rcv}|^2)T_{\rm NS}}
    \label{eq:q-with-nw}
\end{equation}
where\footnote{These same quantities can be expressed in various ways. We have provided a simple direct comparison between this formulation and that of \citet{Roque2021} in App. \ref{app:calibration-comparison}.}
\begin{align}
    \epsilon_{\rm src} &= (1 - |\Gamma_{\rm src}|^2) F_{\rm src}^2 (1-L_{\rm src}), \\ 
    \+{K} &= \big\{|\Gamma_{\rm src}|^2 |F_{\rm src}|^2 , |\Gamma_{\rm src}| |F_{\rm src}| \cos \alpha_{\rm src}, |\Gamma_{\rm src}| |F_{\rm src}| \sin \alpha_{\rm src} \big\} \\
    \+T_{\rm NW} &= \{T_{\rm unc}, T_{\rm cos}, T_{\rm sin}\} \nonumber.
\end{align}
Eq.~\ref{eq:q-with-nw} is the equivalent of Eq. \ref{eq:q-basic} with the inclusion of the complicating factors of the internal reflections and different electrical pathways of the input source compared to the internal loads.
Indeed, following Eq. \ref{eq:TscaToff} we can write:
\begin{align}
    T^{\rm src}_{\rm sca} &= \epsilon_{\rm src}^{-1} (1-|\Gamma_{\rm rcv}|^2) T'_{\rm NS} \\
    T^{\rm src}_{\rm off} &= \epsilon_{\rm src}^{-1} \left[(1 - |\Gamma_{\rm rcv}|^2) T'_{\rm L} - L_{\rm src}T_{\rm amb} - \+{K}\cdot \+{T}_{\rm NW}\right].
\end{align}

\subsection{Solving for the Calibration Temperatures}
\label{sec:calibration:nw-solutions}
Assuming that we can precisely measure $\Gamma_{\rm rcv}$ and $\Gamma_{\rm src}$ (c.f. \S\ref{sec:calibration:s11s}), we are left with five unknown temperatures that are properties of the receiver system rather than the input source: $\{T'_{\rm NS}, T'_{\rm L}, T_{\rm unc}, T_{\rm cos}, T_{\rm sin}\}$.
We will call these the `receiver calibration temperatures', and they must be estimated in order to calibrate any data. To estimate these temperatures, we construct a number of well-characterized `calibration sources' for which we can directly measure the true temperature $T_{\rm src}$, and simultaneously solve for the five temperature functions by considering the measurements from all of these sources.
The choice of the particular calibration sources to use is important -- typically robust solutions require a mixture of sources with very low reflection coefficients (which are suitable for determination of the internal load temperatures $T'_{\rm L}$ and $T'_{\rm NS}$), and sources with larger reflections whose complex reflection coefficients oscillate through several periods over the relevant range of frequencies.
Optimal selection of calibration sources can be aided by optimally sampling their reflection coefficients on a Smith Chart \citep[e.g.][]{Roque2025}, and in principle it is possible to use arbitrarily many calibration sources \citep{Roque2021,DeLeraAcedo2022}.

EDGES uses four calibration sources: two with very low reflection coefficients---one a load at ambient temperature and the other a heated load connected via a short semi-rigid cable---and two with larger reflection coefficients. 
The two sources with larger $S_{11}$'s use the same long cable, which is either short- or open-circuited to yield reflections that are out-of-phase with each other.

With enough independent input calibration sources it would be possible to solve for the five unknown temperatures independently for each frequency channel.
However, it is common to assume that the temperatures evolve smoothly with frequency, so that they can be modeled by low-order flexible linear models, typically polynomials, and thus significantly reduce the number of free parameters. 

Naively, the problem sets itself up as a simple $\chi^2$-minimization. 
Given coefficients of each linear model, $\vec{a} = \{\vec{a}_{\rm NS}, \vec{a}_{\rm L}, \vec{a}_{\rm unc}, \vec{a}_{\rm cos}, \vec{a}_{\rm sin}\}$, we must simply minimize the objective function
\begin{equation}
    f(\vec{a}) = \sum_{\nu,{\rm src}} \frac{\left[T_{{\rm src},\nu} - T_{\rm sca}^{\rm src}(\vec{a}_{\rm NS})Q_{{\rm src}, \nu} - T_{\rm off}^{\rm src}(\vec{a}_{\rm L}, \vec{a}_{\rm unc}, \vec{a}_{\rm cos}, \vec{a}_{\rm sin})\right]^2}{\sigma_{{\rm src},\nu}^2}.
\end{equation}
However, there are many subtleties that must be taken into account.

The first of these is that, as noted in \citet{Murray2022b}, some of the sources may be more sensitive to errors in the measurement of $\Gamma_{\rm src}$ than others, so that their expected systematic error is higher.
This is true of the cable measurements (open and short) for EDGES' calibration. In this case (and in the absence of a trusted model to account for the systematics), to mitigate the systematic effects of these measurements, it is advantageous to use an iterative scheme \citep{Monsalve2017} whereby the internal load temperatures $T_{\rm NS}'$ and $T'_{\rm L}$ are fit only to the non-cable measurements (ambient and hot), which solutions are then used to determine the noise-wave temperatures given the cable measurements, and so on until convergence is reached.

\begin{figure}
    \centering
    \includegraphics[width=\linewidth]{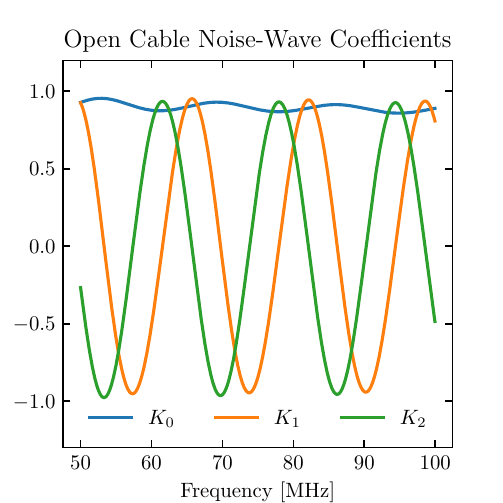}
    \caption{The $K$ coefficients for the open cable calibration source. The sinusoidal period of each is inversely proportional to the cable length (8\,m in the case of this figure, which shows calibration measurements for EDGES-2).}
    \label{fig:Kterms}
\end{figure}

In \ecode, the particular method of solving for the calibration temperatures is flexible: any method that accepts the input measurements and returns an estimated set of calibration temperatures can be defined and used. Here, we will describe some of the details of the iterative scheme, since it is used in the production of the results of \bowman.
The general iterative scheme is outlined in \S3.3 of \citet{Monsalve2017}.
Note that there, instead of the internal load temperatures $T_{\rm NS}'$ and $T_{\rm L}'$, the coefficients are represented as $C_1$ and $C_2$. 
These are simply a scale and offset, respectively, of an initial guess of the internal load temperatures:
\begin{align}
    T'_{\rm NS} = C_1 T^{\rm guess}_{\rm NS} \\
    T'_{\rm L} = T^{\rm guess}_{\rm L} - C_2.
\end{align}
The guesses themselves, as well as the values of $C_1$ and $C_2$, are not physically important and only affect the convergence rate of the algorithm. A key aspect of the algorithm missing from the description in \citet{Monsalve2017} is that the fit of the noise-wave temperatures to the cable data may also account for a single cable delay.
That is, writing the model of the noise-wave contribution to the temperature as
\begin{equation}
    T_{\rm NW} = K_0T_{\rm unc} + K_1 T_{\rm cos} + K_2 T_{\rm sin},
\end{equation}
we instead fit the following model:
\begin{align}
    \label{eq:tnw_model}
    T_{\rm NW} = &K_0T_{\rm unc} \\ \nonumber
    &+ \left[K_1 \cos(2\pi \tau \nu) - K_2 \sin(2\pi \tau\nu) \right] T_{\rm cos} \\ \nonumber
    &+ \left[K_1 \sin(2\pi \tau \nu) + K_2 \cos(2\pi \tau\nu) \right] T_{\rm sin}.
\end{align}
This expression is fit to the residual of the true temperature to the current estimate of the calibrated spectrum without noise-wave terms.
Since $K_1$ and $K_2$ are typically sinusoidal with a phase difference of $\sim90^\circ$ (c.f. Fig. \ref{fig:Kterms}), incorporating a small delay in the model can improve its quality.
The best-fit value of $\tau$ is computed by fitting Eq. \ref{eq:tnw_model} for the calibration temperature coefficients, conditional on a given $\tau$, and minimizing the RMS of the fit residuals over a regular grid of $\tau$.

There are several choices that can be made in the context of the iterative scheme:
\begin{enumerate}
    \item The number of iterations to use and whether to exit when convergence is detected, or to continue for the maximum number of iterations.
    \item The forms of the linear models used for each of the calibration temperatures, as well as the number of terms each uses, and the transformation used on the coordinates (c.f. \S\ref{sec:modeling:models}).
    Currently we allow only the \textsc{Poly} model (Eq. \ref{eq:models:poly}) with $q=0$ and $\mathcal{T}(\nu) = \nu/\nu_{\rm center}$. For $C_1$ and $C_2$ the spacing $p$ can be set, and the number of terms for both must be the same, $n=c_{\rm terms}$. 
    For the noise-wave temperatures, the spacing $p=0$, and the number of terms for all three must be the same, $n=w_{\rm terms}$.
    \item Whether the internal load parameters $C_1$ and $C_2$ should be computed as \textit{models} during the iterations, or instead simply estimated on a per-channel basis and only smoothed with a model after convergence.
    \item Whether the hot load loss (c.f. \S\ref{sec:calibration:source-loss:hotload}) should be forward-modeled onto the `true' temperature $T_{\rm src}$ (as is formulated in Eq. \ref{eq:psrc_full}), or applied in reverse to correct the estimated calibrated spectrum $\hat{T}_{{\rm hot},j} = T_{{\rm sca},j}^{\rm hot}Q_{\rm hot} + T_{\rm off}^{{\rm hot},j}$ on each iteration $j$.
\end{enumerate}

\begin{figure*}
    \centering
    \includegraphics[width=\linewidth]{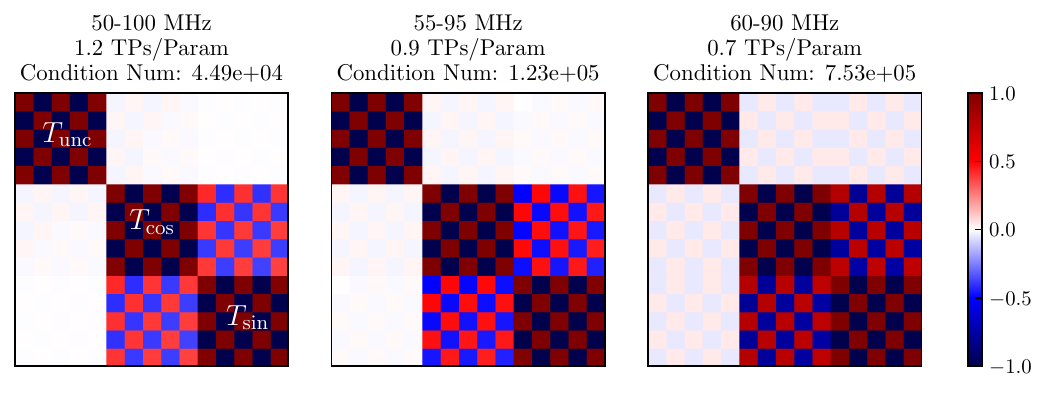}
    \caption{Correlation matrices for 5-term polynomial coefficients of the noise-wave temperatures for EDGES-2 calibration data. From left to right, each correlation matrix considers a smaller frequency range of data, and therefore a different number of turning-points in the reflections per parameter (TPs/Param listed in the title).
    The $T_{\rm cos}$ and $T_{\rm sin}$ models become highly correlated below 1 turning point per parameter, indicated by the off-diagonal blocks evolving from lighter to darker red/blue.. 
    }
    \label{fig:condition-number}
\end{figure*}

An aspect of this iterative algorithm that must be carefully considered is that the amount of structure that can be fit in the noise-wave temperature models is directly related to the amount of structure present in the $K$-coefficients. 
This is because $T_{\rm unc}$, $T_{\rm cos}$, and $T_{\rm sin}$ are modeled with same basis functions, and can only be separated by their products with $K_0$, $K_1$ and $K_2$.
Lack of structure in the $K$ factors results in degeneracies between the parameters of different noise-wave temperature models. 
Since the $K$ are sinusoidal, their structure can be quantified by the number of turning points within the frequency range considered (c.f. Fig. \ref{fig:Kterms}).
This interplay can be seen in Fig. \ref{fig:condition-number}, where we plot the correlation matrices for the parameters of the noise-wave temperatures. 
In this plot, the data are those used in the calibration of \bowman\ (c.f. \S\ref{sec:demo}), and the model for each noise-wave temperature is a polynomial (c.f. \S\ref{sec:demo:workflow:rcvcal}) with 5 terms. 
Each panel shows the correlations after fitting a different range of frequencies centered on 75\,MHz, from a width of 50\,MHz down to 30\,MHz. 
As the frequency range is reduced (left to right panels), the correlation between $T_{\rm cos}$ and $T_{\rm sin}$ parameters (off-diagonal blocks) increases (from light to dark red/blue), as does the condition number\footnote{Defined to be the maximum ratio of the relative error in the best-fit parameters to the relative error in the data.} of the linear system. 
This result means that the residuals become insensitive to changes in $T_{\rm cos}$ and $T_{\rm sin}$. 
However, when these changes get applied to a different set of $K$ coefficients than those used in the fit itself (e.g. the antenna instead of the calibrator sources) they may cause a bias in the calibration.
This bias can be largely avoided by keeping the number of turning points in $K_1$ and $K_2$ per model coefficient above unity. This requirement, in turn, sets a limit on the spectral complexity of the noise-wave temperatures that can be calibrated.

\subsection{Reflection Coefficients}
\label{sec:calibration:s11s}
The model of the expected power at the output of the receiver given an input temperature $T_{\rm src}$, as quantified in Eq. \ref{eq:q-with-nw}, requires knowledge of the complex-valued reflection coefficients of both the input source and the receiver itself. 
Importantly, Eq. \ref{eq:q-with-nw} is formulated assuming that these quantities are defined such that the reference plane of measurement is a 50-$\Omega$ impedance standard at the output of the source, just prior to the SMA1 connector (c.f. Fig. \ref{fig:edges2-diagram}).
Furthermore, $\Gamma_{\rm rcv}$ is defined as the reflection of current coming from the direction of the input source and reflecting back towards it.
Conversely, $\Gamma_{\rm src}$ is defined as the reflection of current coming from the direction of the receiver towards the reference plane and reflecting back towards the receiver.
While the magnitude $|\Gamma|$ is the same regardless of the direction of the reflections, the phase is offset by $180^\circ$ between definitions.
In this section we describe how these quantities are measured and modeled.


\begin{figure}
    \centering
    \includegraphics[width=\linewidth,trim=0cm 1.5cm 0cm 1cm,clip]{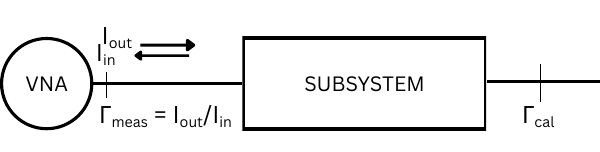}
    \caption{Schematic of the measurement of a VNA, and its calibration. A VNA measures the ratio of current that it injects to the returned current. To calibrate the measurement to a different reference plane (beyond some arbitrary `subsystem') requires application of Eq. \ref{eq:gamma-de-embed} following Eq. \ref{eq:sol-linear-eq}.}
    \label{fig:vna-schematic}
\end{figure}

\subsubsection{Calibrating Reflection Coefficient Reference Planes}
\label{sec:calibration:s11s:refplane}
The reflection coefficient of a given electrical subsystem can be measured by a VNA. The VNA injects a known current into the system and measures the returned current at the port (c.f. Fig. \ref{fig:vna-schematic}). VNA measurements are very accurate, though even small inaccuracies can propagate to biased signal recovery \citep{Sun2024}. 
It is often desirable to know the reflection coefficient at a different point in the system, for example at the other end of a `subsystem', as illustrated in Fig. \ref{fig:vna-schematic}.
Indeed, each VNA must be physically connected to the network, and this connection itself is a subsystem for which the reflection coefficient on the side furthest from the VNA is different than that measured by the VNA at its port.
The process of accounting for this connection is sometimes referred to as `calibrating' the VNA.
More generally, we will refer to the process of determining the reflection coefficient at a point beyond a subsytem given the measurements at the VNA port as `de-embedding' the subsystem.

An accurate formalism for achieving this calibration and de-embedding is presented in \citet{Gonzalez1997} and \citet{Monsalve2016b}.
This formalism utilizes a relation that de-embeds the reflection coefficient based on the scattering parameters of the subsystem:
\begin{equation}
    \Gamma_{\rm cal} = \frac{\Gamma_{\rm meas} - S_{11}}{S_{22}(\Gamma_{\rm meas} - S_{11}) + S_{12}S_{21}}.
    \label{eq:gamma-de-embed}
\end{equation}

The scattering parameters, $S_{11}, S_{12}, S_{21}, S_{22}$, are determined by attaching a set of three `standards'---an Open, Short and Load (OSL) forming a VNA `calibration kit'---to the output of the subsystem.  A key innovation of the EDGES system was to integrate this calibration directly into the receiver as close to the reference plane as possible.  The reflection coefficient is measured with each standard attached and an accurate model of the intrinsic reflection coefficient of each standard is calculated.  The scattering parameters are then determined by solving the following linear equation:
\citep[][Eq. 3]{Monsalve2016b}
\begin{equation}
    \begin{bmatrix}
        1 & \Gamma_O & \Gamma_O \Gamma'_O \\
        1 & \Gamma_S & \Gamma_S \Gamma'_S \\
        1 & \Gamma_L & \Gamma_L \Gamma'_L \\
    \end{bmatrix}
    \begin{bmatrix}
        S_{11} \\ S_{12}S_{21} - S_{11}S_{22} \\ S_{22}
    \end{bmatrix}
    = 
    \begin{bmatrix}
      \Gamma_O' \\
      \Gamma_S' \\
      \Gamma_L' \\
    \end{bmatrix}.
    \label{eq:sol-linear-eq}
\end{equation}
Here the un-primed quantities are the intrinsic reflection coefficients of each standard, computed using the model defined in the Appendix of \citet{Monsalve2016b}, while the primed quantities are the measurements with each standard attached to the subsystem. We solve Eq. \ref{eq:sol-linear-eq} for each measured frequency channel independently. As an implementation detail, the system is solved using the linear least-squares method provided in \texttt{numpy.linalg}.

Calibration of the full EDGES receiver system is essentially comprised of repeated applications of this `de-embedding' to strategic VNA measurements that bring the reference plane to the input at SMA1 (and then application of the noise-wave formalism of \S\ref{sec:calibration:nw-solutions}).
We now define how this de-embedding is applied in practice to measure the reflection coefficients of the receiver and the input sources.

\subsubsection{Measuring $\Gamma_{\rm rcv}$}
\label{sec:calibration:s11s:receiver}
$\Gamma_{\rm rcv}$ is the reflection coefficient defined at the reference plane looking into the receiver input when the receiver is connected to the rest of the signal chain electronics beyond the output SMA3. We measure $\Gamma_{\rm rcv}$ in the lab by externally connecting a VNA to the receiver input (SMA1) and switching the SP2T$_1$ and SP2T$_2$ onto the blue route (Fig. \ref{fig:edges2-diagram}). The VNA measurement is calibrated at the receiver input using external, accurately modeled OSL standards. Specifically, we use the Agilent 85033E calibration kit, which has known values for the impedance, delay, and loss of the `offset' within each standard, as well as reactance of the termination elements. Importantly, while `load' calibration standards nominally have a resistance of 50$\Omega$, our receiver calibration is highly sensitive to this value, and we measure it independently for each new set of calibration measurements using the four-wire method and a high-accuracy (0.01\,$\Omega$), high-resolution multimeter (such as the Fluke\,8845A) \citep{Monsalve2016b}. In \texttt{edges.cal} we define a \texttt{Calkit} type and several instances of calibration kits (and their physical properties) used for EDGES receivers.

\subsubsection{Measuring $\Gamma_{\rm src}$}
\label{sec:calibration:s11s:source}
The reflection coefficient of all input sources (including both calibration sources and the antenna itself) must also be measured and calibrated. 
Recall that $\Gamma_{\rm src}$ is defined at the reference plane of the \textit{input to the receiver} (Fig. \ref{fig:edges2-diagram}), and that it refers to the ratio of current flowing towards the input source from a 50 ohm generator, and reflected back towards the generator.

Obtaining $\Gamma_{\rm src}$ calibrated to the correct reference plane involves measuring the reflection coefficient with a VNA at SMA2 (c.f. Fig. \ref{fig:edges2-diagram}), and then a two-step process of moving the reference plane back to correct location: firstly the `SMA2 + SP4T' subsystem $\overline{r_2r_3}$\footnote{This description assumes that the VNA port has been previously calibrated to 50 ohms, but if this is not the case, any departures from 50 ohms would implicitly be absorbed by the `SMA2 + SP4T' subsystem.} is de-embedded, and then the `SP2T$_2$ + SMA1' subsystem, $\overline{r_1r_2}$. 
These two steps require different configurations of the hardware that must be setup manually.

The steps are as follows:
\begin{enumerate}
    \item \textbf{Setup 1:} Connect calibrator source to SMA1 and VNA to SMA2, ensuring SP2T$_1$ is switched to the `calibration' route (pink network in Fig. \ref{fig:edges2-diagram}).
    \begin{enumerate}
        \item Switch SP4T to the input source and measure $\Gamma^{r_3}_{\rm src}$ (i.e the reflection coefficient of the input source at reference plane $r_3$). 
        \item Switch SP4T to each internal OSL standard in turn, measuring $\Gamma^{r_3}_{\rm OSL, int}$. 
        \item Use Eq. \ref{eq:sol-linear-eq} with $\Gamma'_{\rm OSL} \rightarrow \Gamma^{r_3}_{\rm OSL, int}$ and $\Gamma_{\rm OSL} \rightarrow \{1, -1, 0\}$ to solve for the $S$-parameters of the subsystem $\overline{r_3r_2}$: $S_{\overline{r_2r_3}}$\footnote{Using ideal/generic values of $\{1,-1,0\}$ for the `intrinsic' $\Gamma_{\rm OSL}$ is permitted since we will calibrate this inaccuracy away in Step 2.}.
        \item Use Eq. \ref{eq:gamma-de-embed} to compute $\Gamma_{\rm cal} = \Gamma^{r_2}_{\rm src}$, given $S_{\overline{r_2r_3}}$ and $\Gamma_{\rm meas} \rightarrow \Gamma^{r_3}_{\rm src}$.
    \end{enumerate}
    \item \textbf{Setup 2:} Replace calibrator source at SMA1 with three external calibration standards (OSL) in turn. 
    \begin{enumerate}
        \item Re-measure $\Gamma^{r_3}_{\rm OSL, int}$. This should be done once--it does not depend on the termination at SMA1.
        \item Connect the external OSL calibrators each in turn to SMA1, and for each measure $\Gamma^{r_3}_{\rm OSL, ext}$.
        \item Compute $S_{\overline{r_2r_3}}$ using the same method as in Setup 1 but with the re-measured $\Gamma_{\rm OSL,int}^{r_3}$ from Setup 2 (a).
        \item Use Eq. \ref{eq:gamma-de-embed} to de-embed $\overline{r_2r_3}$ from each external calkit standard in turn, i.e. to compute $\Gamma^{r_2}_{\rm OSL, ext}$ using $S_{\overline{r_2r_3}}$ and $\Gamma_{\rm meas} \rightarrow \Gamma^{r_3}_{\rm OSL, ext}$.
        \item Use Eq. \ref{eq:sol-linear-eq} with the model $\Gamma_{\rm OSL}$ derived from physical properties of the external calkit\footnote{At this point the inaccuracy from Setup 1, in which ideal/generic values were used for the intrinsic reflection coefficients of the internal standards, is calibrated out.} and $\Gamma'_{\rm OSL} \rightarrow \Gamma^{r_2}_{\rm OSL, ext}$ to compute $S_{\overline{r_1r_2}}$.
    \end{enumerate}    
\end{enumerate}

Having obtained both $S_{\overline{r_2r_3}}$ and $S_{\overline{r_1r_2}}$, the raw measurement of $\Gamma^{r_3}_{\rm src}$ may be brought to the reference plane $r_1$ via repeated applications of Eq. \ref{eq:gamma-de-embed}, de-embedding both subsystems in turn. In practice, Setup 2 can only be run in the lab, since it requires manually replacing the input source with calibration standards, which cannot be done in the field (at least, not with the EDGES-2 system), while Setup 1 can be performed any time since it relies only on an automated mechanical switch (SP4T). 
We thus compute $S_{\overline{r_1r_2}}$ in the lab, and assume it is stable over time so that it can be applied to calibrate the antenna in the field, given periodic measurements of $\Gamma^{r_3}_{\rm ant}$ and $S_{\overline{r_2r_3}}$. Similarly to our calibration of $\Gamma_{\rm rcv}$ (Sec. \ref{sec:calibration:s11s:receiver}), for Setup 2 of the calibration of $\Gamma_{\rm src}$ we use the Agilent 85033E calibration kit as the external OSL. In order to keep the system as stable as possible, we use active thermal control in the field.

\subsection{Source Losses}
\label{sec:calibration:source-loss}
Sources connected indirectly to the receiver port, for instance via a cable, in general experience some loss. For passive devices, this loss means that their effective noise temperature is not equal to their physical temperature. This effect needs to be taken into account. The loss of noise power coming from an input source over a particular two-port network (e.g. a cable, balun or connection component) can be computed as \citep{Memo132,Monsalve2017}:
\begin{equation}
    1 - L_{\rm src} = \frac{|S_{12}S_{21}|(1 - |\Gamma_{\rm in}|^2)}{(1 - |\Gamma_{\rm out}|^2)|1 - S_{22}\Gamma_{\rm out}|^2},
    \label{eq:cable-loss}
\end{equation}
where the $S$-parameters are those of the two-port network itself,
$\Gamma_{\rm in}$ is the reflection coefficient at the reference plane of the interface between the input source and the two-port network, and $\Gamma_{\rm out}$ is the reflection coefficent at the reference plane of the output of the two-port network. 

Since the reference plane for calibration reflection coefficients is at the input to the receiver, this is equivalent to $\Gamma_{\rm out}$ for input sources (whether they are calibration sources or the antenna itself).
Conversely, $\Gamma_{\rm in}$ can be computed from $\Gamma_{\rm out}$ and the $S$-parameters of the two-port network by de-embedding with Eq. \ref{eq:gamma-de-embed}.

The loss $L_{\rm src}$ has both a multiplicative and additive effect on the temperature passing through a component, according to \citet{Monsalve2017}:
\begin{equation}
    T_{\rm out} =  (1 - L_{\rm src}) T_{\rm in} + T_{\rm amb} L_{\rm src}.
\end{equation}
That is, in the high-level framework of \S\ref{sec:overview} (Eq. \ref{eq:high-level-data-reduce}), we have $\mathbf{G}_{\rm antloss} = \mathbf{I}(1 - L_{\rm src})$ and $T_{\rm loss} = T_{\rm amb}L_{\rm src}$, where $T_{\rm amb}$ is the physical temperature of the two-port network over which the loss is experienced (often this is the ambient temperature of the observation).

\subsubsection{Hot Load Loss}
\label{sec:calibration:source-loss:hotload}

In the EDGES-2 calibration setup, only the hot load is indirectly connected to the receiver input, via a short semi-rigid cable. 
Its loss, $L_{\rm hot}$, is computed with Eq. \ref{eq:cable-loss} using $\Gamma_{\rm out} \equiv \Gamma_{\rm hot}$ (as measured and calibrated to the reference plane using the formalism of \S\ref{sec:calibration:s11s:source}) and $S$-parameters that are measured directly with a VNA in the lab (using the formalism of \S\ref{sec:calibration:s11s:refplane} and \citet{Monsalve2016b}). 
The $S$-parameters of this semi-rigid cable are generally \textit{not} measured for each calibration, and we instead use a single measurement for all re-calibrations.

\subsubsection{Antenna Loss}
\label{sec:calibration:source-loss:antloss}
The EDGES-2 antenna is connected to the receiver input via both a Roberts balun and a short connector. Similarly to the hot-load loss, we use Eq. \ref{eq:cable-loss} with $\Gamma_{\rm out} \equiv \Gamma_{\rm ant}$. Here, however, instead of directly measuring the $S$-parameters we use an analytical model based on transmission line theory \citep{Roberts1957}. In this model, each component (e.g. the balun and the connector) is treated as a coaxial cable consisting of inner and outer conductors as well as a dielectric, where their physical dimensions and electrical constitutive parameters are known. This information can be used to compute electrical properties of the cable, which can be used in standard transmission line theory \citep{ramo1953fields}
to compute scattering parameters\citep[e.g.][]{Allen2026}, and the components are cascaded to produce the full scattering matrix (c.f App. \ref{sec:cascading}).
For more details on this approach see the EDGES memo series \citep{Rogers2012a,Rogers2015a}


Presenting the full equations relating physical dimensions and electrical constitutive parameters of coaxial cables to their electrical properties, and these electrical properties to the scattering parameters, is beyond the scope of this work, and we refer the reader to works such as \citet{ramo1953fields,dunsmore2020handbook} for these well-known results.
However, we note that the \texttt{edges.cal.ee} module contains convenient classes for defining coaxial cables and their properties, representing them as transmission lines, and computing scattering matrices. 
The module also contains methods for converting the representation of scattering matrices (e.g. between $\mathbf{T}$ and $\mathbf{S}$, c.f. App. \ref{sec:cascading}) and combining multiple components that are connected either in parallel, series or cascaded.

\subsection{Beam-Factor Chromaticity Correction}
\label{sec:calibration:bfcc}
The antennas sensitivity pattern, or beam, changes with frequency, and this chromaticity translates into spectral structure in the antenna temperature since this beam weights the foregrounds.
This effect has been extensively studied in the literature \citep{Bernardi2015,Mozdzen2017,Tauscher2018,Anstey2021,Hibbard2020,Hibbard2024,Mozdzen2019,Sims2023,Sims2025} and is difficult to account for.
A common strategy to mitigate the chromatic effects of the beam is to divide by a `beam factor chromaticity correction' factor \citep{Monsalve2017a,Mozdzen2019,Sims2023}.

This strategy can only provide a perfect correction of the data when employing perfect beam and sky foreground models. Nonetheless, it tends to \textit{reduce} the chromaticity of the observations so long as the beam model is sufficiently accurate.

The beam factor as a function of frequency at some local sidereal time (LST) $\alpha$ can be defined as \citep{Mozdzen2019}
\begin{equation}
    G_{\rm beam}(\nu, \alpha) = \frac{\int_{2\pi} d^2 \Omega\ T_{\rm sky}(\Omega, \nu_{\rm ref}, \alpha) B(\Omega, \nu)}{\int_{2\pi} d^2 \Omega\ T_{\rm sky}(\Omega, \nu_{\rm ref}, \alpha) B(\Omega, \nu_{\rm ref})},
    \label{eq:beamfac}
\end{equation}
or alternatively with the same form but with $\nu_{\rm ref} \rightarrow \nu$ for the factor $T_{\rm sky}$ (e.g. Eq. 4 from \citealt{Monsalve2017a} and Eq. A1 from \citealt{Sims2023}).
Under either definition, $G_{\rm beam}$ is by construction unity at $\nu_{\rm ref}$.

In practice, computing Eq. \ref{eq:beamfac} requires making a number of choices.
The first choice is the sky model, $T_{\rm sky}$. 
In \ecode, we provide a unified interface to several publicly available all-sky models, including that of \citet{Haslam1982} and de-striped versions of it \citet{Remazeilles2015} at 408\,MHz, as well as maps at lower frequencies, such as those of \citet{Landecker1970} at 150\,MHz and \citet{Guzman2011} at 45\,MHz.
Maps at higher frequencies are also provided that enable studies of Radio Recombination Lines \citep[RRLs][]{Vydula2024}, e.g. the WHAM \citep{Haffner2003,Paradis2012}, Planck CO maps \citep{Ade2014} at 115\,GHz and HI4PI \citep{Bekhti2016} at 1.4\,GHz. 
All of these maps are defined at a single frequency.
To determine the sky model over the range of frequencies measured by EDGES, we provide a number of spectral index models.
These models specify the spatial evolution of the spectral index, and include simple models such as a uniform spectral index and others that evolve smoothly as a function of galactic latitude.

A second choice is the beam model, $B$. 
In \ecode, we provide a unified interface to simulated beam data (e.g. from Feko and CST) similar to that provided by \texttt{pyuvdata}, but without features unnecessary for the purposes of global-signal data.
To perform the product of beam with sky, the beam must be interpolated to the positions of the pixels of the sky model. 
While there are many subtleties of beam interpolation for complex-valued antenna voltage beams \citep{Kittiwisit2025}, the real-valued antenna power beam needed here poses fewer challenges.
We allow either spherical bivariate spline interpolation\footnote{Using the \texttt{RectSphereBivariateSpline} method from \texttt{scipy}} or regular-grid interpolation in which the zenith is represented multiple times in the grid. 
Differences in the interpolated values between these methods are insignificant.

A third choice concerns how to compute the beam factor over a \textit{range} of LSTs corresponding to the observations, rather than at a single LST.
To do this, \ecode\ computes the numerator and denominator separately for a regular grid of LSTs covering the observed range.
The resolution of this grid can be set by the analyst.
Following this choice, both the numerator and denominator can be interpolated to a different set of LSTs (potentially at a higher resolution).
Then, two basic options are available: either (i) average $G_{\rm beam}(t)$ over all LSTs within the observed range or (ii) average the numerator and denominator separately within the observed LST range, and then form their ratio. 
In addition, to ensure the beam factor is spectrally smooth in case of numerical artifacts, it may be fit with a smooth linear model and the model can be used instead of the directly simulated values.
Ideally, this should be avoided by ensuring that the simulation itself is accurate. 

\section{Flagging} \label{sec:flagging}
All spectral data are contaminated to some level by various instrumental or foreground artifacts.
While effects that broadly affect many channels and times (e.g. beam chromaticity, receiver gains, galactic foregrounds) must be well-modelled and calibrated, effects that are compact in time and/or frequency (e.g. terrestrial RFI, lightning, solar events, hardware malfunction) may simply be flagged.
This flagging has the bonus that these effects need not be well-modeled in principle\footnote{There is a caveat to this picture; while channels and integrations can be flagged wholesale, in order to achieve unbiased averages requires `in-painting' the flagged gaps (c.f. \S\ref{sec:averaging}), which can be viewed as a kind of crude modeling-and-subtraction of the systematic.}; flags can be set on a relative basis (i.e. flagging outliers), or through inspection of ancillary data (e.g. ambient humidity or environmental temperature fluctuations). 

In general, given the extreme precision requirements on global 21\,cm measurements, any data suspected of being contaminated beyond the ability to accurately calibrate (to within a few mK on scales of tens of MHz) should be removed.
There are two negative consequences of flagging that must be kept in mind.
The first is a loss of sensitivity. However, given that the requisite sensitivity can be reached in a matter of a few days of observations with most experiments, this is not a strong argument for foregoing suspect data.
The second is that it is possible to bias the measurements depending on the location of the flags in frequency and time (see \S\ref{sec:averaging} for a discussion of this effect). The severity of this problem is reduced when there is no correlation between the metric defining the flags and the amplitude/shape of the foregrounds and cosmic signal. Nevertheless, to achieve unbiased results after flagging does require accurate in-painting of the flagged regions, and large contiguous flagged regions can be difficult to model, potentially resulting in spurious structure.

To minimize over-flagging, a common strategy is to flag in stages, flagging the small amount of worst-offending raw data, then averaging to higher signal-to-noise and finding new outliers, and continuing in this fashion until the data are suspected to be clean. 
For EDGES data, which are single-polarization, we will always talk about data being two-dimensional, referring to time and frequency.
At any stage of averaging, either one or both of these axes may have been smoothed and down-sampled. 
Furthermore, the `time' axis may really represent different chunks of time, or it may represent bins of LST into which multiple raw integrations have been averaged (potentially from multiple days of observation).

For EDGES data, there are two main kinds of flagging strategies: the first is to flag entire spectra---either an entire integration, an entire LST bin, or an entire night, but always \textit{all channels} for that particular time selection. This strategy can be further broken down into flags based on external measurements (e.g. weather) and absolute or relative characteristics of the spectral data themselves.
The second kind of strategy flags individual integration-channels (i.e. arbitrary pixels on a 2D time-frequency `waterfall'). 
This strategy can be done by considering each integration (or LST-bin) independently or by considering the entire 2D waterfall simultaneously. 
We refer to this frequency-dependent flagging as `RFI flagging', though the integration-channels flagged may have contributions from effects other than terrestrial RFI. 
All flagging in \texttt{edges-analysis} is defined in the \texttt{edges.filters} module.

\subsection{Integration-Based Flagging Strategies}
\label{sec:flagging:integrations}
\subsubsection{Ancillary/Metadata Flags}
\label{sec:flagging:integrations:auxiliary}
The following filters based on external data are available within \texttt{edges.filters}:
\begin{itemize}
    \item \textbf{Time-Dependent Auxiliary Data} (\texttt{aux\_filter}): Since \gsd\ objects contain a time-length table attribute \texttt{auxiliary\_measurements} which may contain arbitrary quantities, the data can be flagged based on threshold ranges on any particular auxiliary quantity contained here. For EDGES-2, parallel measurements of ambient humidity, ambient temperature, the receiver physical temperature and the uncalibrated power spectral density (PSD) measured by the Analog-to-Digital Converter (ADC) are recorded and can be associated with each dataset. While any of these may be used to filter particular integrations, only the ADC PSD is used in the specific analysis demonstrated in \S\ref{sec:demo}.
    
    \item \textbf{Sun Altitude} (\texttt{sun\_filter}): flag data when the sun is within a certain set of altitudes (e.g. above the horizon). The sun is a bright \citep[$>10^5$\,K;][]{ZhangZhangPei2022} and variable radio source with bursts that complicate spectral model fitting \citep{Vasanth2025}.
    
    \item \textbf{Moon Altitude} (\texttt{moon\_filter}): flag data when the moon is within a certain set of altitudes (e.g. above the horizon). The moon is reflective and can be a source of reflected terrestrial RFI \citep{Pattison2026}.

    \item \textbf{Astronomical Source Altitude} (\texttt{sky\_coord\_filter}): flag data when a particular fixed coordinate (in RA/Dec) is within a certain set of altitudes. For simplicity, we use Astropy's \texttt{from\_name} functionality so that particular named objects can be excluded from the data.

    \item \textbf{Galactic Centre Altitude} (\texttt{galaxy\_filter}): as a special case of the sky-coordinate filter, we provide a convenience function that flags data when the galactic centre is at given elevations (default: above the horizon).
\end{itemize}

\subsubsection{Flags based on Spectral Properties}
\label{sec:flagging:integrations:spectrum}
Sometimes a particular single spectrum (whether raw integration or LST bin) has properties which indicate that the entire spectrum should be ignored. 
In \texttt{edges-analysis} the following strategies are defined for removing entire spectra based on their internal properties (note that here we give general filters, while in \S\ref{sec:demo:workflow:singleday} we will give specific examples of these filters in practice):
\begin{enumerate}
    \item \textbf{Negative Power} (\texttt{negative\_power\_filter}): Power measurements $P_{\rm src}$ should be strictly positive. Any negative values indicate a malfunction of the spectrometer and therefore these integrations should be flagged.
    \item \textbf{Peak Power} (\texttt{peak\_power\_filter)}: When RFI spikes are too large there is a danger of the system being saturated, resulting in non-linearities that affect the entire spectrum. A strategy to detect this case is to compute the ratio of the highest single-channel power within some frequency range to the mean within some (potentially different) frequency range: $R = P_{\rm max}/ \langle{P}\rangle_{\{\nu_{\rm min}, \nu_{\rm max}\}}$ and flag the integration if it is beyond some large threshold. To make the strategy more robust, the mean ignores any channels whose power is $>10\%$ of $P_{\rm max}$. This is a crude filter, but if the threshold is large, it finds the worst offenders.
    \item \textbf{Strong RFI Spike} (\texttt{single\_channel\_spike\_filter}): this filter again focuses on spikes that may cause non-linearity,  but uses a different strategy that targets single channels. 
    Here, if any channel in a given band is more than some threshold greater than the average of its immediately neighbouring channels, the entire integration is flagged. This is applied to the FM band specifically in the analysis of \S\ref{sec:demo}.
    \item \textbf{Model Residual RMS} (\texttt{rms\_filter}): some systematics are not compact spikes like RFI, but rather manifest in broader spectral structures. When these are beyond the expectation of any of the calibration models, it may be best to remove the data. This removal can be done by fitting a smooth model within some band and computing the RMS of the residuals to this model, finally flagging if this RMS is beyond a given threshold. 
    This strategy can be applied early in the pipeline with very simple rigid models, targeted frequency ranges, and large thresholds.
    Later in the pipeline it can be applied with more flexible models, on larger frequency ranges, and lower thresholds.
    \item \textbf{Fractional Power in Band} (\texttt{power\_percent\_filter)}: this test computes the ratio of the sum of all power in a particular frequency range to the sum of all power within the entire spectrum, and flags the integration if this ratio is outside a certain threshold range (either too small or large). 
    Such a filter can be used in highly bespoke ways by tuning the threshold empirically; for example in the EDGES-2 pipeline it is used to identify rare cases of hardware failure.
\end{enumerate}

\subsection{RFI Flagging} \label{sec:flagging:rfi}
All methods of RFI flagging centre around identifying particular integration-channels (or contiguous groups of such) as being statistically significant outliers. 
Many algorithms for RFI identification have been developed for low-frequency radio astronomy, typically in the context of interferometric observations (e.g. SSINS \citep{Wilensky2019a}, AOFlagger \citep{Offringa2012}, and the HERA xRFI algorithms \citep{HERA22}).
Common elements of all algorithms are to recognize that RFI is generally compact in both time and frequency, that there are particular frequency bands where it is more likely to be present (e.g. FM and Orbcomm).
Since the pattern of RFI in a 2D time-frequency waterfall is difficult to model, and its amplitude is all but impossible to predict, most strategies utilize some form of outlier detection, rather than an absolute criterion.
This approach requires establishing a baseline both for the expected underlying signal without RFI, and its level of channel-to-channel variation (from e.g. thermal noise). 
Differences between approaches are typically centered on how they determine these two baselines.

The primary challenge is that to establish the baseline model of non-RFI-contaminated data (and its variance) requires using the data themselves. 
Large RFI excursions in the data can significantly bias these baseline estimates, resulting in poor flagging accuracy. The \texttt{edges.filters.xrfi} module provides several RFI-detection algorithms that utilize different strategies for mitigating this bias.
Having multiple algorithms available as plug-and-play replacements means that the ramifications of the choice of algorithm can more easily be explored.
At this time, \texttt{edges.filters.xrfi} only provides algorithms that treat each time-sample independently, though in the future this implementation will be expanded to consider full 2D waterfalls simultaneously.

\subsubsection{A Generalized Iterative RFI Algorithm}
\label{sec:flagging:rfi:general-iterative}
One way to mitigate the effect of large RFI excursions is to flag iteratively. In this way, the worst offenders can be flagged first, so that they do not contribute to the determination of the baselines in future integrations.
In this section, we outline a general iterative algorithm that we will fill out later.

We have as input $\vec{d}$ representing the single-integration spectrum data, $\vec{w}$ representing the weights of each datum (zero when it is already flagged from some other filter) and $\vec{\xi}_{\rm init}$ representing a set of a priori flags that will be re-considered after the first iteration, but which are useful for ignoring data that are likely to bias the establishment of the initial baselines. 
We also give a minimum number of iterations required before returning, $n_{\rm min}$. Alg. \ref{alg:general-rfi} outlines the algorithm.

\begin{algorithm}
\begin{algorithmic}
    \State $\vec{\xi}_{\rm out} \gets \vec{\xi}_{\rm init}$
    \State $i \gets 0$
    \State $done \gets \False$
    \While{$\neg done \AND i < n_{\rm max}$}
        \State $\vec{w}_i \gets \vec{w} \cdot \vec{\xi}_{\rm out}$
        \State ${\rm threshold} \gets \Call{SetThreshold}{i}$
        \State $\vec{p}_i \gets \Call{SetModelParams}{i}$
        \State $\vec{m} \gets \Call{ComputeDataModel}{\vec{d}, \vec{w}_i, \vec{p}_i}$
    
        \State $\vec{p'}_i \gets \Call{SetSTDModelParams}{i}$
        \State $\vec{\sigma} \gets \Call{ComputeSTDModel}{\vec{d},\vec{m}, \vec{w}_i,\vec{p}'_i}$
    
        \State $\vec{z} \gets (\vec{d} - \vec{m})/\vec{\sigma}$
        \State $\xi_i \gets z > {\rm threshold}$
        \State $\xi_{\rm i} = \Call{Watershed}{\xi_i}$
        \If{$\xi_i = \xi_{\rm out} \AND \vec{p}_i = \vec{p}_{i-1} \AND \vec{p'}_i = \vec{p}'_{i-1}$}
            \State $done \gets \True$
        \EndIf
        \State $\xi_{\rm out} \gets \xi_i$
        \State $i \gets i+ 1$
    \EndWhile
\end{algorithmic}
\caption{General iterative algorithm for flagging RFI.}
\label{alg:general-rfi}
\end{algorithm}

In Alg. \ref{alg:general-rfi}, there are many generic methods that must be specified. Features of the algorithm are that on each iteration, the method used for determining the model of the true data (i.e. without RFI) and the model of the true standard deviation can be (optionally) updated, by changing the method parameters (indicated by \textsc{SetModelParams} and \textsc{SetSTDModelParams}). Once these are determined, flags are computed simply by comparing the $Z$-score of the residuals to a threshold, which may also be adapted on each iteration, and then applying the \textsc{Watershed} algorithm, which we define below. 
The flags are fully reconsidered on each iteration; flags from the previous iteration are used only to inform the baseline models, and may be un-flagged depending on these updated baselines.
The algorithm stops when no new flags are found and the model parameters have not changed between iterations, or when $n_{\rm max}$ iterations have occurred. 

Before describing various possible choices for the methods to compute the data and variance models, we describe the \textsc{Watershed} algorithm. 
This algorithm simply recognizes that when RFI is detected in a particular channel, it is more likely that the surrounding channels might also be contaminated, even if they were not beyond the set threshold.
To this end, the \textsc{Watershed} algorithm accepts a list of pairs $(\kappa'_j, n_j)$ where $\kappa_j \in \mathbb{R} \geq 1$ and $n_j \in \mathbb{N}$ and flags the $n_j$ channels to the left and $n_j$ channels to the right of each channel in which $z/{\rm thresh} > \kappa_j$.

Computing the baseline models of the data and their variance can be done in numerous ways. In general, the data model can be computed by applying a low-pass filter to the spectrum.
For simplicity and ease of propagating errors, in \texttt{edges.filters.xrfi} we only support linear filters, such that the filter may be expressed as an $N_\nu \times N_\nu$ matrix $\bf{W}$, and the model of some data $\vec{y}$ expressed as $\mathbf{W}\vec{y}$.
In practice, \texttt{edges.filters.xrfi} supports building $\bf{W}$ in three ways: (i) as a circulant matrix defined by any compact convolution kernel that is normalized to unity (e.g. Gaussian, BoxCar; (ii) as a general linear model (utilising \texttt{edges.modeling}, c.f. \S\ref{sec:modeling}), $\bf{W} = \bf{M}(\bf{M}^T\bf{M})^{-1}\bf{M}^T$; and (iii) as a median filter\footnote{While the median filter can be written as a matrix $\bf{W}$ containing only zeros and ones, it is technically not a linear operation since the matrix itself depends on the data vector.}.
For smooth parametric models (e.g. polynomial or Fourier-based models), a critical choice is the number of terms to fit\footnote{For simplicity, we consider only linear models in \texttt{edges-analysis}.}.
Analogously, for low-pass filters, the size of the kernel determines the spectral complexity of the resulting smoothed data. 
In either case, the hyper-parameters of $\bf{W}$ (i.e. the kernel size or number of linear coefficients) may be updated between iterations via the \textsc{SetModelParams} procedure.
Ultimately, we may write the smoothed data as
\begin{equation}
    \+m = \bf{W} \+d.
\end{equation}

The model for $\vec{\sigma}$ can be generically computed in much the same manner as $\vec{m}$, but considering the `data' to be the squared residuals, $r_i = (d_i - m_i)^2$.
Since the variance is positive definite, it is advantageous to apply $W$ on the \textit{logarithm} of the squared residuals, $\rho_i = \ln r_i$, restoring the final result to linear scaling, and adjusting for the Jacobian of the transformation:
\begin{equation}
    \+\sigma = 1.888\times \sqrt{\exp(\bf{W} \+\rho)}.
\end{equation}
For the median-filter, these transformations are not optimal, and instead we use the median absolute deviation (MAD):
\begin{equation}
    \+\sigma_{\rm MAD} = \sqrt{0.456 \times \bf{W}_{\rm med} \+r }.
\end{equation}

In \texttt{edges.filter.xrfi}, $\bf{W}$ can be specified differently for computing $\+d$ versus $\+\sigma$.
In practice, convolutional kernels (which estimate the running mean of the spectrum) tend to perform poorly when modeling $\+d$ since it is often intrinsically a steep power-law. 
Centred symmetric kernels produce biased estimates of the mean in such scenarios.

\subsubsection{Sliding Window Flagging}
\label{sec:flagging:rfi:sliding}
Here we describe a small alteration to the general iterative RFI algorithm that is important because it is the method used to produce the results in \bowman. 
In this algorithm, instead of computing the full $\vec{\sigma}$ for all channels, and then using it to determine the flags based on $z$, we compute $\sigma_\nu$ and its corresponding $z_\nu$ and flags for the central channel of a sliding window.
Importantly, if the channel is flagged, its weight is zero for the next sliding of the window. 

This approach is fundamentally different from Alg. \ref{alg:general-rfi}, and has some undesirable properties. For example, the resulting flags depend on which direction the window slides across the channels. 
The difference between the algorithms can be summarized by requiring that all lines of Alg. \ref{alg:general-rfi} between the computation of $\sigma$ and the application of the \textsc{Watershed} algorithm be inside a loop over frequency channels, where on each iteration of this inner loop, only a single central channel can be flagged, plus any that are flagged by the watershed algorithm.

\section{Averaging} \label{sec:averaging}
Once contaminated data have been flagged, it is often desirable to average the data to increase the SNR, which aids in identifying lower-level contamination and reducing the ultimate data volume.

EDGES data have two axes over which they can be averaged: times and frequencies. However, it is conceptually useful to differentiate between different observing days at the same (or a similar) LST vs the same night and different LSTs. Information is preserved when averaging is performed over identically distributed data (or close thereto). 
The only axis over which this is generally the case is \textit{days}. 
Nevertheless, since EDGES does not observe on a regular LST grid, each night will observe slightly different LSTs. Thus, averaging within frequency bins or LST bins is not information-preserving---the signal and foregrounds evolve within the bin. 
While the effect of this convolution kernel on the cosmic signal is very minor (as it evolves slowly over both frequency and LST), a more important problem arises when there are frequency-dependent flags.

As is well-studied in the interferometry literature, frequency-dependent flagging patterns can cause spectral structure when spectra with varying amplitudes are averaged together \citep{Wilensky2022,Chen2025}.
The cause is very simple: two different spectra averaged together will generally result in the mean of the two, but if one is flagged for a particular channel, the average will jump suddenly to the value of the unflagged spectrum. If the expectation of the two spectra is the same, this is not a problem. However, in practice, these two spectra may be drawn from different days in the same LST bin. However, one may be towards the start of the bin, and the other towards the end (perhaps they are separated by an hour). Over an hour, the beam-weighted amplitude of the sky can change significantly, which would result in large spurious spectral structure.
Indeed, \citet{Lewis2021} showed that for flagging fractions of $\sim 10\%$, the RMS error on the naively-averaged spectrum within LST bins at the native EDGES resolution of just 39\,sec is expected to be more than 50\,mK unless the Galaxy is below the horizon. This effect is not only a problem when data are flagged.
The same arguments apply when the data are non-uniformly weighted \citep{Murray2020c}.
A common solution to this problem is to `in-paint' flagged data with an estimate of its true value. 
We can extend this strategy to arbitrary non-uniformly weighted data.
Let $m$ be a model of the data $d$, and assume that the data weights are inverse variances (i.e. the weighted-average is minimum-variance), then the standard weighted average is:
\begin{equation}
    \bar{d} = \mathcal{W}^{-1} \sum_i w_i d_i,
\end{equation}
which has the variance 
\begin{equation}
    {\rm Var}(\bar{d}) = \mathcal{W}^{-1} \equiv \left(\sum_i w_i\right)^{-1}.
\end{equation}
Nevertheless, as discussed above, if $d_i$ is drawn from a different distribution than $d_j$, and $w_i \neq w_j$ this can lead to spectral structure being introduced. 
Instead, we use the model:
\begin{equation}
    \hat{\bar{d}} = N^{-1}\sum_i^N m_i + \mathcal{W}^{-1} \sum_i^N  w_i (d_i - m_i).
    \label{eq:unbiased-average}
\end{equation}
That is, we use the un-weighted sum of the models plus the weighted sum of residuals. In the case that all data have the same mean and the models are good representations of the data, the expectation of $\hat{\bar{d}}$ is the true mean. Furthermore, under these same assumptions, the variance of this average is still $\mathcal{W}^{-1}$. Indeed, if all weights are the same, then this reduces to a standard mean. If the weights are binary -- i.e. all zeros or ones -- then this is equivalent to in-painting the `missing' data and taking a standard average. 

In \texttt{edges-analysis} every function that averages spectra---whether over frequency channels, LSTs or days---is able to use this formula.
This is achieved by \texttt{GSData} objects being able to hold both \texttt{.data} and \texttt{.residuals} arrays (the latter can be added by particular functions that model the data).

A choice must also be made about how to propagate the expected thermal noise of the data. In \texttt{pygsdata} (as in \texttt{pyuvdata}), the `variance' of the data is propagated simply by tracking the number of raw samples that have contributed to each particular datum. 
Under the assumption of Gaussianity of the spectra, this number of samples can be translated into a variance via the radiometric equation for thermal noise. 
In detail, this can yield biased estimates of the variance for data averaged from heterogeneous sources, since the true resulting variance is the sum of the variances of each source, which may vary not only due to their number of samples, but also due to the temperature of the sky for each particular datum. On the other hand, in general raw data are not calibrated, and therefore knowing the true temperature of the sky before averaging is often difficult or impossible. Simply propagating the number of samples is a reasonable compromise. 

While rigorous propagation of the number of samples is the preferred strategy, for legacy reasons \ecode\ provides a flexible set of strategies for choosing the weights to use in both averaging and modeling. 
In brief, the code supports setting the weights to be (i) the flagged-samples $\xi_i n_i$, (ii) binary flags, $\xi_i$, (iii) binary flagged-samples $w_i = 1\  {\rm if}\ \xi_i n_i>0\ {\rm else}\ 0$, (iv) un-flagged samples $n_i$ and (v) uniformly unity. These choices are consistently applied across all averaging and modeling functions.
In all averaging functions, the averaged number of samples (which can be used to compute the variance of the average) is computed as the sum of the flagged-samples $\sum_i \xi_i n_i$ regardless of the weighting strategy. 

With these general concepts introduced, we now describe some particulars concerning averaging over frequencies and LSTs (averaging over days is a simple matter of applying Eq. \ref{eq:unbiased-average} where the dataset includes each night). 
Frequencies and LSTs are different because typically these are \textit{binned} rather than completely averaged. 
Binning is simply \textit{down-sampling}, which is a well-studied concept of signal-processing. 
To avoid aliasing effects when down-sampling a signal requires first smoothing with a convolutional kernel and then decimating by keeping every $k^{th}$ sample such that the samples are spaced by the desired bin size \citep{Harris2004}.
The choice of kernel should be driven by considerations of the factor by which the signal is to be downsampled. 

In \ecode, down-sampling frequencies can use either the box-car or Gaussian kernels for smoothing before decimation. In the former case, the size of the box-car is set equal to the decimation factor, in order to avoid correlations between neighbouring points in the resulting signal.
This choice is equivalent to performing a direct weighted-average (using Eq. \ref{eq:unbiased-average}) over all channels in each bin.
The Gaussian kernel is defined to have a FWHM of $1.2\Delta$, where $\Delta$ is the decimation factor. 
This kernel introduces correlations between resulting averaged channels, which are currently unaccounted in \ecode's propagation of uncertainties.
Nevertheless, when the decimation factor is $\geq 8$ this correlation is less than 0.1\%.
After smoothing, some channels that were originally flagged (or had zero samples) become un-flagged as they average in surrounding channels.
We provide an option to re-flag these channels based on the criteria of either how many immediately adjacent channels were originally flagged, or the number of samples in the pixel after convolution with the same Gaussian kernel. 

Down-sampling in LST is somewhat complicated by the periodic nature of LSTs, as well as the potential for entire LSTs to be missing from the data (and therefore the signal be sampled irregularly). 
To overcome this complication, we choose a specific LST at which the phase wraps (the lower edge of the first bin), and we only use direct application of Eq. \ref{eq:unbiased-average} within each LST bin (equivalent to a box-car kernel, if the data are regular).

\section{Application: Reproducing Results of B18} 
\label{sec:demo}
As a demonstration of the data interface and analysis algorithms presented in this work, we reproduce the results of \bowman.
This application of the software achieves a few purposes; 
(i) a demonstration of how these algorithms can be used;
(ii) a more comprehensive and explicit description of the precise analysis choices that formed this important result compared to the original publication; and 
(iii) a reproducible and publicly accessible workflow that can be used by the community to more thoroughly examine the result.

\begin{figure*}
    \centering
    \resizebox{\textwidth}{!}{%
\definecolor{colorout}{HTML}{4A90E2}

\tikzstyle{line} = [draw, -latex']

\tikzstyle{input} = [rectangle, draw, fill=red!40, 
    text width=7em, text centered, minimum height=2em]

\tikzstyle{inputhalf} = [rectangle, draw, fill=red!40, 
    text width=5em, text centered, minimum height=2em]

\tikzstyle{product} = [circle, draw, fill=colorout, 
    text width=5em, text centered, minimum height=2em, text=white]

\tikzstyle{calibration} = [rectangle, draw, fill=blue!40, 
    text width=10em, text centered, rounded corners, minimum height=2em]
\tikzstyle{averaging} = [rectangle, draw, fill=green!40, 
    text width=10em, text centered, rounded corners, minimum height=2em]
\tikzstyle{modeling} = [rectangle, draw, fill=purple!40, 
    text width=10em, text centered, rounded corners, minimum height=2em]
\tikzstyle{filter} = [rectangle, draw, fill=orange!30, 
    text width=10em, text centered, rounded corners, minimum height=2em]
\tikzstyle{computecal} = [rectangle, draw, fill=pink!60, 
    text width=10em, text centered, rounded corners, minimum height=2em]

\begin{tikzpicture}[
    node distance = 0.5em, 
    auto,
    arrow/.style={-{Stealth[scale=1.2]}, thick}
]
    \node [input] (raw) {Raw Spectra\\Night 1};
    \node [input,below=of raw] (raw2) {Raw Spectra\\Night 2};
    \node [input,below=of raw2] (raw3) {Raw Spectra\\Night 3};
    \node[below=0.5] (dots) at (raw3.south) {\Huge ...} ;
    
    \node [filter,right=2.5em of raw] (lst_select) {Galactic Centre Filter};
    \node [filter,below=of lst_select] (adc) {Auxiliary data filter};
    \node [filter,below=of adc] (pp) {Power Percent filter};
    \node [calibration,below=of pp] (dcal) {Dicke Calibration};
    \node [calibration,below=of dcal] (tcal) {Scale to approximate temperature};
    \node [filter,below=of tcal] (orb) {Orbcomm filter};
    \node [filter,below=of orb] (fm) {MaxFM filter};
    \node [filter,below=of fm] (rms) {RMS filter};
    \node [averaging,below=of rms] (lst_bin) {Average Integrations};
    \node [filter,below=of lst_bin] (selectfreq) {Select Frequencies $40-100$\,MHz};
    \node [filter,below=of selectfreq] (xrfi) {RFI Excision (I)};
    \node [averaging,below=of xrfi] (gauss_smooth) {Frequency Downsample (x8)};
    
    \node[shape=rectangle, draw, fit=(lst_select) (gauss_smooth), inner sep=4] (dayavgnb) {};

    \draw[arrow] (raw.east) -- (raw.east -| dayavgnb.west);
    \draw[arrow] (raw2.east) -- (raw2.east -| dayavgnb.west);
    \draw[arrow] (raw3.east) -- (raw3.east -| dayavgnb.west);

    \node [input,right=2.5em of pp] (rawcaldata) {Lab-Based Calibration Data};

    \node [computecal,right=2.0em of rawcaldata] (rcvcal) {Compute $\vec{T}_{\rm cal}$};
    
    \node[shape=rectangle, draw, fit=(rcvcal) (rcvcal), inner sep=4] (rcvcalnb) {};

    \draw[arrow] (rawcaldata.east) -- (rcvcal.west);

    \node [input,below=1.5em of rawcaldata] (ants11data) {Antenna VNA Measurements};

    \node [computecal,right=2.0em of ants11data] (ants11cal) {Calibrate $\Gamma_{\rm ant}$};
    \node [computecal,below=of ants11cal] (ants11model) {Model  $\Gamma_{\rm ant}$};

    \node[shape=rectangle, draw, fit=(ants11cal) (ants11model), inner sep=4] (ants11nb) {};

    \draw[arrow] (ants11data.east) -- (ants11cal.west);

    \node [computecal,below=1.5em of ants11nb] (bccf) {Compute $G_{\rm BFCC}$};    

    \node [input,below=3.1em of ants11data] (beamdata) {FEKO Beam};
    \node [input,below=of beamdata] (skydata) {HASLAM Sky};

    \node[shape=rectangle, draw, fit=(bccf) (bccf), inner sep=4] (bccfnb) {};

    \draw[arrow] (beamdata.east) -| ($(bccf.west) - (0.3, 0)$) |- (bccf.west);
    \draw[arrow] (skydata.east) -| ($(bccf.west) - (0.3, 0)$) |- (bccf.west);
    
    
    \node [calibration,right=5em of rcvcal] (cal) {Receiver Cal.};
    \node [filter,above=of cal] (freq50) {Select Freq. 51-99\,MHz};
    \node [calibration,below=of cal] (loss) {Balun+Connector Loss Correction};
    \node [calibration,below=of loss] (bf) {Beam Chromaticity Correction};
    \node [averaging,below=of bf] (gauss_2) {Frequency Downsample (x8)};

    \node[shape=rectangle, draw, fit=(freq50) (gauss_2), inner sep=4] (datacalnb) {};

    \draw[arrow] (rcvcalnb.east) -- (cal.west);
    \draw[-,style=thick] (ants11nb.east) -| ($(cal.west) - (0.75, 0)$);

    \draw[arrow] (bccfnb.east) -| ($(bf.west) - (0.75, 0)$) |- (bf.west);

    \draw[arrow] ($(dayavgnb.north east) + (0, -0.1)$) -| ($(datacalnb.north) + (0.2, 0.1)$) -| ($(datacalnb.north) + (0.2, 0)$);
    \draw[arrow] ($(dayavgnb.north east) + (0, -0.3)$) -| ($(datacalnb.north) + (0.0, 0.1)$) -| ($(datacalnb.north) + (0.0, 0)$);
    \draw[arrow] ($(dayavgnb.north east) + (0, -0.5)$) -| ($(datacalnb.north) + (-0.2, 0.1)$) -| ($(datacalnb.north) + (-0.2, 0)$);
    
    \node [filter,below=3.5em of datacalnb] (objrms) {Per-night RMS filter};
    \node [filter,below=of objrms] (xrfi2) {RFI Excision (II)};
    \node [averaging,below=of xrfi2] (day) {Average over nights};
    
    \node[shape=rectangle, draw, fit=(objrms) (day), inner sep=4] (nightavgnb) {};
     
    \draw[arrow] (datacalnb.south) -- (nightavgnb.north);
    \draw[-,style=thick] ($(datacalnb.south) + (0.2, 0)$) |- ($(nightavgnb.north) + (0, 0.5)$);
    \draw[-,style=thick] ($(datacalnb.south) - (0.2, 0)$) |- ($(nightavgnb.north) + (0, 0.5)$);

   \node [product, left=2.5em of nightavgnb] (final) {Averaged Spectrum};
    \draw[arrow] (nightavgnb.west) -- (final.east);

    \node[computecal,text width=7em, left=2.5em of gauss_smooth] (compute-legend) {CAL SETUP};
    \node[calibration, text width=7em,above=of compute-legend] (cal-legend) {CALIBRATION};
    \node[averaging, text width=7em, above=of cal-legend] (avg-legend) {AVERAGING};
    \node[filter, text width=7em, above=of avg-legend] (filter-legend) {FILTER};
    \node[input, text width=7em, above=of filter-legend] (input-legend) {INPUT DATA};
    
    \node[shape=rectangle, draw, thick,fit=(input-legend) (compute-legend), inner sep=6] (legbox) {};
    
    \node[above=0.2] (leg) at (legbox.north) {\huge LEGEND} ;
    
\end{tikzpicture}
    }
    \caption{The full analysis pipeline used to produce case `H2' of Fig. 1 of \bowman. 
    Each colored box is a specific analysis step, while the black rectangles that contain multiple steps are major components of the pipeline, each implemented as a single Jupyter notebook.
    The flow is generally from left to right, and from top to bottom within each major component.
    Different colors represent different kinds of analysis steps (see figure legend).}
    \label{fig:b18-pipeline}
\end{figure*}
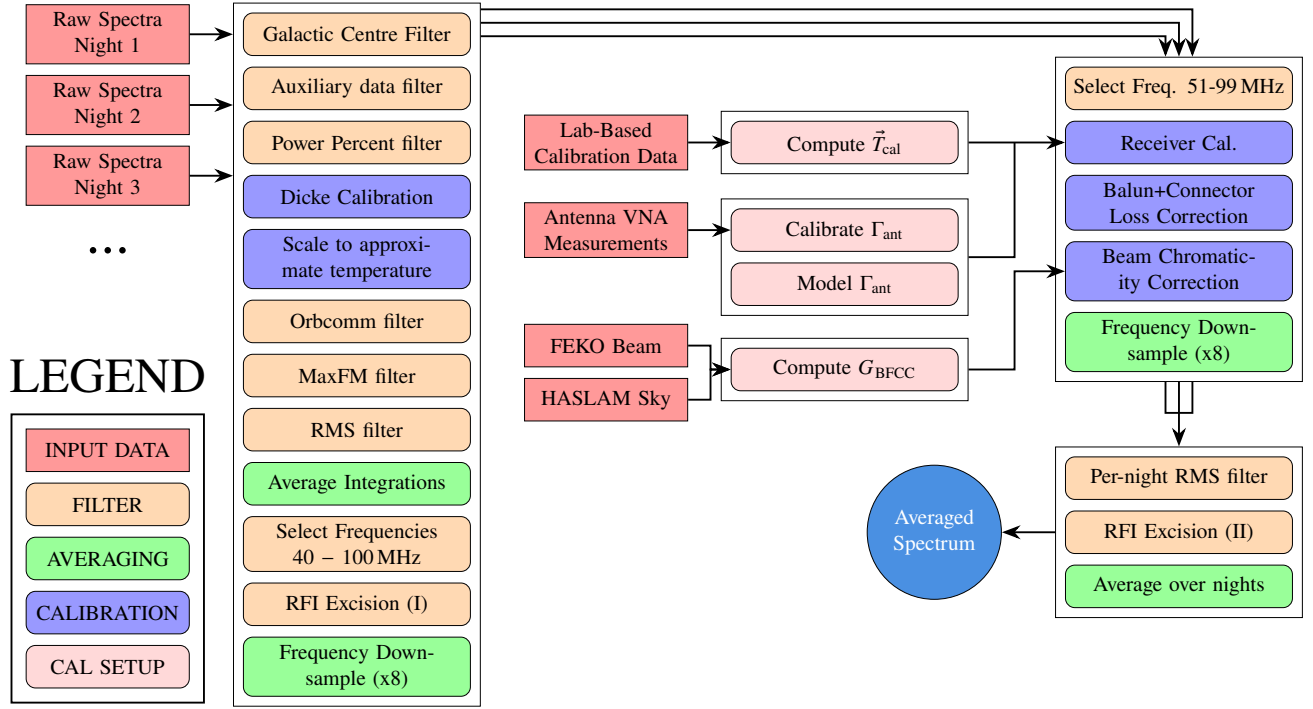

Fig. \ref{fig:b18-pipeline} outlines the analysis pipeline designed for this dataset. Starting with 138 days of raw spectrum data (from September 2016 through to April 2017), a full set of lab-based calibration data, and a set of measurements of the reflection coefficients of the antenna taken in the field, the data first undergoes quality checks that flag entire integrations, then is averaged over full days before a first round of RFI excision is performed. 
Following this, the data is calibrated both for the receiver gain and also other losses such as beam chromaticity.
Once calibrated, the averaged spectra from each night are averaged together, after some final deeper quality checks, and a final round of RFI excision is performed.

In this paper, our goal is simply to detail the analysis choices that \textit{were made}, with some brief description of their motivation. 
In particular, we will \textit{not} investigate the impact of modifying these choices, except in some specific cases.
Neither is it the purpose of this work to defend these choices.
These important goals will be pursued in \mahesh, using the flexible pipeline we demonstrate here.

Note that there are many other possible pipeline pathways that could have been chosen, beyond the choice of analysis parameters at each step. 
For instance, the data could be first binned in narrow LST bins, and quality checks could be performed by comparing data in the same LST bin over days. 
Alternatively, instead of a two-step averaging (over integrations in a night, then over days), a multi-step process could be envisaged (narrow LST bins, then days, then gradually grow the size of the LST bins) in which between each averaging further quality checks are applied.
Since the analysis methods in \ecode\ are loosely coupled, resting on the unified data interface of \gscode, it makes no strong assumptions about the flow of analysis, and such diverse pipelines are easy to implement.

\subsection{Pipeline Framework}
\label{sec:demo:nextflow}
While \ecode\ does not place any strong constraints on how an analysis of data may be implemented (e.g. whether the data is interactively explored in a Jupyter notebook, more rigidly processed through a python script, or through any other data-processing pipeline framework), we have found that for reproducible analysis, using a dedicated data-pipeline framework is highly advantageous.
Such pipelines, which are becoming very popular due to the rise of machine-learning applications, enable simple and robust parallelization of the compute, reproducible and transparent runs over different parameter choices, the ability to resume processing after a failure, and many other benefits.

In this work, we use the popular \texttt{nextflow} framework \citep{DiTommaso2017}\footnote{\url{https://nextflow.io}}, which has been developed primarily for science applications (rather than machine-learning or data science applications) and is used by several 21\,cm experiments \citep[e.g. NenuFAR;][]{Munshi2024}.
In conjunction with \text{nextflow} as a workflow orchestration tool,
we implement each major component of the pipeline as an executable Jupyter notebook using the \texttt{papermill} library to enable parameterization of each notebook.
This approach, which follows the lead of recent work by the HERA experiment \citep{HERA2026}, is helpful for interactive debugging and data exploration (e.g. when a notebook fails for a particular piece of data, the partially-executed failed notebook can be loaded interactively and the data leading to the failure can be directly explored, accelerating development time). 
It is also helpful for including inline data visualization and well-formatted in-context documentation explaining the choices made in the format of a dynamically-produced report. 

We make the \texttt{nextflow} pipeline definition and associated executable notebooks (as well as resulting executed notebooks/reports) for this work publicly available on Github\footnote{\url{https://github.com/edges-collab/edges-bowman2018-pipeline}} \citep{steven_murray_2025_17526742}.
This pipeline is fully able to be run on any system---the required datasets will be automatically downloaded and cached upon first run. 
We also make the data directly available via Zenodo; both the data required to solve for the receiver calibration \citep{Murray2025_EDGESCalZenodo}\footnote{\url{https://doi.org/10.5281/zenodo.18091240}} and the spectra from the field \footnote{\url{https://loco.lab.asu.edu/edges/edges-data-release/}}. 

\subsection{Workflow} \label{sec:demo:workflow}
We now describe the full analysis pipeline, specifying each methodological choice and each parameter.
Each subsection will describe a single major step of the pipeline that is implemented as a single executable notebook (represented by an outer black rectangle surrounding individual analysis components in Fig. \ref{fig:b18-pipeline}). 
For a concise summary of the pipeline and choices at each step, see Appendix \ref{app:pipeline-table}.

\subsubsection{Single-Day Flagging and Averaging}
\label{sec:demo:workflow:singleday}
The raw spectrum data (observed in the field) are first read as \gsd\ objects.
These data consist of $N_{{\rm intg}, j}$ integrations on day $j$, 
each of which consists of three spectra---the internal load, internal load + noise source and antenna input---with $N_{\nu}=32768$ channels covering $0-200\,{\rm MHz}$.
There are 138 days recorded in this dataset, with a total of 297,168 integrations (about 1000 hours on-sky). The primary aim of the first major processing step is to remove all integrations that are considered of questionable quality, average the remaining integrations together, and then flag channels containing RFI before down-sampling over frequency.

The first three filters are applied to the raw power spectral densities.
The first filter removes integrations where the galactic centre is above the horizon, ultimately retaining LSTs $23.76-11.76$\,hr: about half of the data (c.f. \S\ref{sec:flagging:integrations:auxiliary}). 
The Galactic centre is very bright in the low-frequency radio band, which intensifies the requirements on calibration precision. Removing these integrations with this filter lowers these requirements.
There is a small subtlety here in that in \gscode\ the calculation that determines the LST from the observation time by default uses the precise methods provided by \texttt{astropy}.
However, the LST-setting method can be changed by the user, and for this analysis we use a more approximate formula that was used in the legacy pipeline. 
Using the more accurate formula from \texttt{astropy} makes only a very small difference to the LSTs, and these LSTs are only used in two places in the pipeline: once here to determine whether the galaxy is above the horizon, and then again when computing the beam factor correction.
While the differences in LST are small, using the more accurate method would mean that a few entire integrations would be either removed or kept at odds with the original pipeline, which changes the absolute value of the final averaged spectrum non-negligibly.
The impact of these small changes will be discussed in \mahesh.

The second filter is an \texttt{aux\_filter} (c.f. \S\ref{sec:flagging:integrations:auxiliary}) that flags integrations for which the analog-to-digital converter (ADC) recorded power is greater than 0.35 V$^2{\rm Hz}^{-1}$. The particular ADC used in EDGES-2 quotes a saturation level of 0.5 V$^2{\rm Hz}^{-1}$, however tone injection tests indicate that values above 0.35 V$^2{\rm Hz}^{-1}$ leak narrow-band spikes across the spectrum to unacceptable levels \citep{Memo307}. Such strong spikes can occur due to lightning \citep{Rogers2020}.

\begin{figure}
    \centering
    \includegraphics[width=\linewidth]{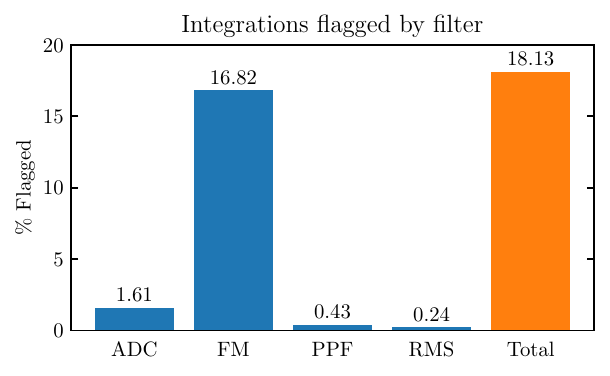}
    \caption{Breakdown of the fraction of data flagged by specific filters that flag entire integrations. Each bar represents a percentage of the data after already being down-selected for the galactic centre. The Orbcomm filter is omitted for clarity as it caused no flags. The total flagged fraction is less than the sum of the individual filters, because some of the flags overlap.}
    \label{fig:flag-breakdown}
\end{figure}

The third filter uses the \texttt{power\_percent\_filter} (c.f. filter (v) in \S\ref{sec:flagging:integrations:spectrum}) 
to flag integrations where the fractional power (from the antenna input) contained in the range $100-200$\,MHz (i.e. the upper half of the spectrum) compared to the entire spectrum is outside the range $0.7-3$\%. 
This range was set empirically, to coarsely filter out data for which the spectrum does not exhibit the broadly correct shape (for instance, assuming Galactic foregrounds that scale as approximately $\nu^{-2.6}$, and a passband that strongly declines below 10\,MHz but is otherwise constant, yields an expected ratio of about 1.6\%---close to the middle of the $0.7-3$\% range).
In practice we find that the 0.16\% of integrations that are \textit{uniquely} flagged by this filter are flagged due to a few channels of extremely bright RFI---typically in the sub-Orbcomm band ($120-135$\,MHz). 
Less than 0.5\% of integrations fall afoul of this filter (Fig. \ref{fig:flag-breakdown}, `PPF'), and it is likely that in future analyses it will be replaced by something more targeted.

The remaining filters require data to be at least approximately calibrated, and so the next step is to perform the Dicke-switch calibration (Eq. \ref{eq:dickecal}), which removes the time-variation from the gains, and then to apply an \textit{approximate} scaling of the resulting data $Q$ to temperature units, using an \textit{a priori} guess for the calibration temperatures $T^{\rm guess}_{\rm sca}=1000$\,K and $T^{\rm guess}_{\rm off}=300$\,K (cf. Eq. \ref{eq:TscaToff}). 

Following this approximate temperature calibration, we apply three more filters that remove entire integrations. The first applies the \texttt{peak\_power\_filter} (filter (ii) of \S\ref{sec:flagging:integrations:spectrum}) to filter integrations excessively affected by Orbcomm RFI. In particular the peak temperature in the Orbcomm range, $137-138$\,MHz, is compared to the mean temperature between $80-200$\,MHz, and the integration is flagged if this ratio is $>40$. This filter does not find any flags in this dataset, and is therefore omitted from Fig. \ref{fig:flag-breakdown}.

The second applies the \texttt{single\_channel\_spike\_filter} (filter (iii) of \S\ref{sec:flagging:integrations:spectrum}) exclusively in the FM range, $88-120$\,MHz with a threshold of 200\,K. This filter removes integrations where strong RFI in the FM range may cause non-linearities over the entire spectrum. The threshold is essentially arbitrary, but catches any strong single-channel spikes. This is the most impactful filter, flagging more than 15\% of data (c.f. `FM' in Fig. \ref{fig:flag-breakdown}).

Finally, integrations are flagged based on an \texttt{rms\_filter} (filter (iv) of \S\ref{sec:flagging:integrations:spectrum}). In this case, the data is first restricted to the range $60-80$\,MHz which is expected to be reasonably free from RFI. The model used as a prior on the spectral shape within this band is a simple power-law, $T_{75} (\nu/75\,{\rm MHz})^{-2.5}$, where the amplitude $T_{75}$ is fit independently to each integration. Integrations with an RMS greater than 200\,K within this range are flagged.

Fig. \ref{fig:flag-breakdown} indicates the fraction of data flagged for each of these per-integration filters, as well as the total fraction flagged (which is less than the sum of the individual filters due to overlap). 
The dominant cause of flagging is very high amplitude single-channel RFI spikes in the FM band.
Overall about 18\% of the integrations are flagged, with more than 16\% coming from extreme FM spikes. 

Note that in this analysis we do not remove either daytime data, nor data for which the moon is up, though these cases have been investigated (e.g. Extended Data Table 1 of \bowman).
The fraction of integrations for which the Galactic centre is down and the sun is up in this dataset is about 53\%, making it a significant aspect to consider.

After removing these integrations, we average together all remaining spectra (in the form of approximate temperatures). 
Since thus far we have only flagged entire integrations, the weights are spectrally-uniform, and thus averaging with the unbiased form Eq. \ref{eq:unbiased-average} is equivalent to a standard (unweighted) average. 
We are left with a single spectrum for each day of observation, at raw frequency resolution.
From this, we select channels between $40-100$\,MHz for further processing and calibration (this band was chosen primarily due to a combination of our bandpass filter, which declines sharply below 40\,MHz and above 120\,MHz, as well as RFI beyond 100\,MHz). 

\begin{figure}
    \centering
    \includegraphics[width=\linewidth]{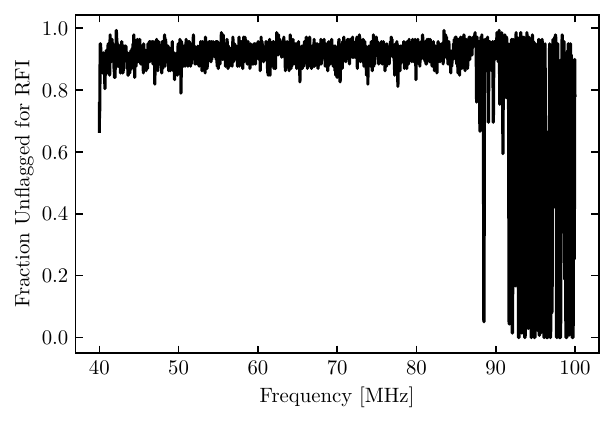}
    \caption{Fraction of data in each channel across the full dataset that is unflagged by the first round of RFI excision.}
    \label{fig:rfi-occupancy}
\end{figure}

We now perform the first channel-dependent filter for RFI. For this, we use the iterative filter presented in \S\ref{sec:flagging:rfi:general-iterative}, with the flags determined for a sliding window, as described in \S\ref{sec:flagging:rfi:sliding}. 
In particular, to compute the data model, we use a \textsc{Fourier} model (Eq. \ref{eq:models:fourier}) with $n=37$ terms, with a coordinate transform $\mathcal{T}(\nu) = (\nu -  40\,{\rm MHz})/90\,{\rm MHz}$.
To compute the model of the standard deviation, we use a box-car filter of size 614 channels ($1/16^{th}$ of the size of the remaining spectrum between $40-100$\,MHz).
We flag at a threshold of $Z > 2.5\sigma$, and use a watershed filter such that channels adjacent a flagged channel are always flagged, the 8 channels on either side are flagged if $Z>25\sigma$, and the 16 channels on either side are flagged if $Z>250\sigma$.
We do not change either the threshold or the models on any of the iterations, and the flagging always converges within 100 iterations.
The fraction of data in each channel across the full dataset that is \textit{not} flagged in this round of RFI excision is shown in Fig. \ref{fig:rfi-occupancy}.
For most of the band, channels are flagged about 10\% of the time, while in the FM band, flagging rates can approach 100\%. 
The total flagging fraction for all channels is 13.5\%.

Note that this is one of the few steps where extremely small numerical differences can have a large impact on the absolute temperature of the final averaged spectrum. 
In particular, due to the high condition number of the 37-term \textsc{Fourier} basis over this range of frequencies ($\sim 1.4\times10^8$), very small numerical differences in the basis functions or the numerical algorithm used for solving the linear system can result in noticeable differences in the best-fit model.
As an example, differences between evaluating cos and sin in C versus \texttt{numpy} result in an RMS deviation in the basis functions computed by each of the order of $10^{-15}$.
Nevertheless, using the \textit{same} code (i.e. the legacy C-code), based on simple QR-decomposition, to compute the best-fit parameters based on these numerical basis functions results in best-fit models whose residual RMS is 13\,mK. 
Increasing the precision of $\pi$ from 10 decimals (used in the C-code) to full double-precision results in an increased residual RMS of 28\,mK. 
In fact, even using the same algorithm but using the dot-product from \texttt{numpy} (wich uses pairwise-summation) rather than a direct summation as done in the legacy C-code with the \textit{exact same} basis functions results in an even larger RMS difference of over 400\,mK. 

While these differences are modest, they are exacerbated by the fact that the model of $\sigma$ is multiplicative in the $Z$-scores, and the fractional differences are rather large. Indeed, the above absolute differences correspond to 0.5\%, 3.5\% and 27\% in $\sigma$ respectively. 
Furthermore, since the algorithm makes a sharp cut based on $Z_{\rm thresh}$, and identified flags propagate to further modeling, the differences can become increasingly impactful as the algorithm progresses.
The three cases mentioned here each result in approximately 100 channels flagged differently, in a spectrum for which about 1300 channels are ultimately flagged---a 7.5\% differential.

We found that the primary culprit here is the QR-decomposition method implemented in the legacy C-code.
This method does not re-normalize the basis functions to reduce the condition number of the system, as most modern methods do (e.g. \texttt{numpy.linalg.lstsq}). 
While using \texttt{lstsq} on the same data instead of the QR-decomposition as implemented in the C-code results in an even larger impact of over 200 channels flagged differently, it is far more stable to slightly changing the basis functions (e.g. changing the precision of $\pi$), resulting in no flag differences.
Additionally, comparing the best-fit models between the QR-decomposition and \texttt{lstsq} shows that the RMS residuals are slightly reduced for the latter. 

While it is possible to exactly reproduce the RFI flags of \bowman\ by directly wrapping the relevant original functions for computing the basis functions and performing the QR-decomposition, this discussion reveals that using the more stable and precise methods from \texttt{numpy} is better, and so we accept the differences that arise from this choice. 
These differences can be surprisingly large; flagging some different channels on a per-night basis (usually channels that are close to the threshold) mostly adds some extra noise-like structure to the absolute differences. 
However, later stages of processing eliminate entire days when their RMS residual to a low-order model exceeds a certain threshold, and this threshold can be exceeded due to the new `noise' from different per-night flags. 
Since each night has a different effective weighting as a function of LST, eliminating an entire night can significantly change the LST distribution for the final averaged spectrum, changing its absolute amplitude by a few K. 
Regardless, since none of these changes are expected to be correlated with the 21\,cm signal, we don't expect the ultimate interpretation to be significantly affected.
Put another way, while the resulting absolute spectrum may differ by a few Kelvin, its residual to a smooth foreground model should not be expected to differ by more than a mK.
This is borne out by Fig. \ref{fig:xrfi_result_compare}, in which we show both the case in which the RFI flags are manually taken directly from the legacy pipeline and injected into the \ecode\ pipeline, and the case in which we use the more robust \texttt{lstsq} model-fitting algorithm with all other choices set to match the legacy pipeline.

Finally after excising the RFI, we down-sample the spectra by a factor of eight channels, using the Gaussian kernel (c.f. \ref{sec:averaging}). 
Here we re-flag each channel that was originally flagged, and also any channel where the number of samples after convolution is less than a quarter of the full complement.

\subsubsection{Receiver Calibration}
\label{sec:demo:workflow:rcvcal}
The next major data-processing step is calibration. However, to perform the calibration
we require the calibration solutions, determined from data taken in the lab.

The data used to determine the calibration solutions for this work were recorded in September 2015. Many of the measurements are VNA recordings of reflection coefficients, particularly of OSL standards.
For this dataset, our external calkit is a Keysight\footnote{Formerly Agilent} 85033E kit. 
Models of the intrinsic $\Gamma_{\rm OSL}$ (required to solve Eq. \ref{eq:sol-linear-eq}) require physical properties of the calkit standards, which we take from the relevant specsheet, with two exceptions: (i) the delay incurred by the `offset' in the Load standard is set as 30\,ps\footnote{Equations relating these properties to the model reflection coefficients can be found in the Appendix of \citet{Monsalve2016b}, and the properties themselves are encoded in \ecode.}, and (ii) the resistance of the Load standard is accurately measured with a high-resolution multimeter independently for each OSL measurement (c.f. \S\ref{sec:calibration:s11s:source}).
The dependence of the reflection coefficients on other physical properties (e.g. capacitance) are significantly weaker than the dependence on the resistance of the Load \citep{Monsalve2017}.

\begin{figure}
    \centering
    \includegraphics[width=\linewidth]{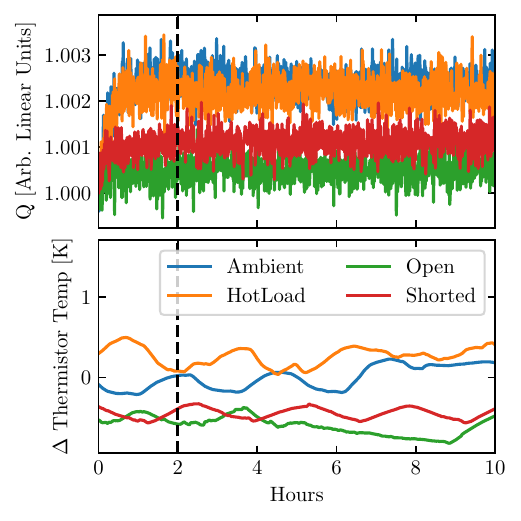}
    \caption{Warmup of the calibration loads over time. We remove the first two hours of data so that the source temperature has stabilized. The top panel shows the PSD ratio, $Q$, normalized so that the first measurement is one. The bottom panel shows the thermistor reading of the temperature of the source, as a difference to the assumed value (399 K for the hot load, and 296 K for the others).}
    \label{fig:warmup-spec}
\end{figure}

In brief, the dataset consists of:
\begin{itemize}
    \item Six sets of four VNA readings to measure $\Gamma_{\rm rcv}$ (c.f. \S\ref{sec:calibration:s11s:receiver}). Each set contains one measurement of the VNA connected to the receiver input, and three direct measurements of the external OSL standards. The Load resistance is measured at 49.98\,$\Omega$ for each set. We average the six calibrated $\Gamma_{\rm rcv}$. These six display $\sim1\%$ systematic (spectrally smooth) variations around the mean value.
    
    \item Three sets of six VNA readings of OSL standards---one OSL internal to the receiver, one external---to compute the $S$-parameters of the `internal switch' (i.e. $\overline{r_1r_2}$; c.f. \S\ref{sec:calibration:s11s:source}). Each set is measured at a different controlled ambient temperature (18.67$^\circ$\,C, 27.16$^\circ$\,C, and 35.31$^\circ$\,C), and the resistance of the external Load standard is measured at 50.13\,$\Omega$, 50.12\,$\Omega$ and 50.11\,$\Omega$ respectively. The final $S$-parameters are linearly interpolated between these measurements to the ambient temperature at which the calibration procedure is performed, measured at 27$^\circ$\,C.

    \item For each calibration source (ambient, hot load, open cable, shorted cable):
    \begin{itemize}
        \item Spectra (2250 integrations for the ambient, 2126 for the hot load, 3295 for the open, 5420 for the shorted cable).
        \item Thermistor readings of the temperature of the input, recorded on a cadence of one second for the duration of the spectrum readings.
        \item Four VNA readings of reflection coefficients: one recorded with the internal four-position switch pointed to the input source, and the other three pointing to the internal OSL standards. These are used to produce a \textit{calibrated} reflection coefficient at the reference plane of the input to the receiver for the input source, following \S\ref{sec:calibration:s11s:source}.
    \end{itemize}
    \item Three VNA measurements of the external OSL attached to the end of the semi-rigid cable that is used to connect the hot load to the receiver input. These are used to compute the $S$-parameters of the cable, which are then used to calculate the loss of the hot load (c.f. \S\ref{sec:calibration:source-loss:hotload}). The resistance of the Load standard is measured at 50.11\,$\Omega$.
\end{itemize}
This full dataset is publicly available\footnote{\url{https://doi.org/10.5281/zenodo.18091240}} \citep{Murray2025_EDGESCalZenodo}. 
A subtlety is that in the public dataset we provide thermistor readings as described, but we do not use these directly to determine the calibration solutions in this work, instead using approximate assumed temperatures for each source, namely 296\,K for the ambient and long-cable sources, and 399\,K for the hot load.
Fig. \ref{fig:warmup-spec} shows that the thermistor measurements are within 1\,K of these assumed values throughout the calibration measurements.
These choices produce a minimal change to the final calibration, which will be investigated in \mahesh.

We first average all Dicke-calibrated spectrum readings from each input source. When averaging the spectra, we ignore the first two hours of recordings from each source as the spectrometer warms up (c.f. Fig. \ref{fig:warmup-spec}). We restrict the spectra to the frequency range $40-110$\,MHz, and after averaging, we down-sample the spectra with the Gaussian kernel by a factor of eight, before further restricting the range to $50-100$\,MHz.

All of the reflection coefficients and $S$-parameters are recorded at a resolution of 250\,kHz by the VNA, which is a different resolution than the spectra. 
To simultaneously re-grid the reflections and to remove measurement noise, we fit smooth linear models to them, evaluating the models at the frequencies of the (down-sampled) spectra\footnote{Note that while this reduces measurement noise in the reflection coefficients used in downstream analysis, it introduces a non-trivial correlated modeling uncertainty that should in principle be propagated through the pipeline, quite apart from any systematic uncertainty in the measurements. This is left for future work.}.
Since the different sources/components have different spectral characteristics, different model complexities are required for each. 
For all reflection coefficients, we use the \textsc{Fourier} model (Eq. \ref{eq:models:fourier}) with a coordinate transform $\mathcal{T}(\nu) = (\nu -  50\,{\rm MHz})/ 75\,{\rm MHz}$, and fit this model to both real and imaginary components separately.
However, the number of terms differs for each; for $\Gamma_{\rm src}$ and the hot load short cable $S$-parameters we use 27 terms, whereas for $\Gamma_{\rm rcv}$ we use 11 terms\footnote{The sensitivity of the analysis to these choices will be investigated in \mahesh.}.
When fitting these models, we first remove a pure tone with a best-fit delay $\tau$, i.e. we fit the linear model to $\Gamma \exp(2\pi \tau \nu)$, and then re-apply the delay to the model when evaluating, $\Gamma_{\rm mdl} = \exp(-2\pi \tau\nu)\mathbf{M}_{\rm fourier}[\vec{a}_{\rm real} + i\vec{a}_{\rm imag}]$.
This single tone is removed directly as there is no a priori guarantee that its wavelength corresponds to a harmonic of the frequency range being modeled, which would otherwise complicate the Fourier model fit.
The delay $\tau$ is fit by maximizing the objective function
\begin{equation}
    {\rm max}_{\tau} \left|\sum_\nu \Gamma_\nu \exp(2\pi i \nu \tau)\right|
\end{equation}
on a regular grid between $\tau\in\{-1, 100\}$\,ns in steps of 0.1\,ns.
This inherently optimises for a delay that keeps the data as coherent as possible.
This is not done for the hot load $S$-parameters.

To solve for the calibration temperatures, we use the iterative method described in \S\ref{sec:calibration:nw-solutions}.
Specifically, we use $c_{\rm terms}=6$ and a spacing $p=0.5$ (c.f. Eq. \ref{eq:models:poly}) for the internal load \textsc{Poly} temperature models, while for the noise-wave temperatures we use $w_{\rm terms}=5$.
Within the iterative loop we \textit{do not} smooth the current estimate of $T_{\rm NS}'$ and $T_{\rm L}'$, and only smooth them at the end of the procedure. 
We apply the hot load loss to the \textit{observed} hot load spectra rather than to the known `true' hot load temperature. 
We run for eight iterations (although the procedure converges in four in this case).

\subsubsection{Calibrating $\Gamma_{\rm ant}$}
\label{sec:demo:workflow:ants11}
The antenna reflection coefficients are calibrated in a similar manner to $\Gamma_{\rm src}$ as described in the previous section.
Measurements of $\Gamma^{r_3}_{\rm ant}$ are taken in the field by switching SPT2T$_1$ to the in-field VNA connected to SMA2 (c.f. pink route in Fig. \ref{fig:edges2-diagram}).
These are calibrated according to the procedure in \S\ref{sec:calibration:s11s:source}, using $S_{\overline{r_1r_2}}$ previously measured in the lab (as described in \S\ref{sec:demo:workflow:rcvcal}).

The VNA used to measure $\Gamma_{\rm ant}^{r_3}$ in the field is not identical to that used in the lab to measure the calibration sources. However, since we calibrate the measurements of each VNA back to a common reference plane (c.f. \S\ref{sec:calibration:s11s:refplane}) this is not in principle a problem (or, it would only be a problem if the VNA itself was mis-calibrated, in which case it would be a problem even if the VNA was the same in the lab and in the field!). 

In order to reduce the amplitude of noise fluctuations on $\Gamma_{\rm ant}$, we take measurements over more than 25 hours at a cadence of 34 seconds. 
Since the correction for the internal switch is temperature-dependent, we simultaneously measure the internal temperature of the receiver via a thermistor\footnote{An Omega ON 930 44006.}, and independently correct each measurement by an $S_{\overline{r_1r_2}}$ that is a weighted average of the two internal switch measurements closest in temperature (recall that we measured the internal switch at 18.76, 27.16 and 35.31 $^\circ$C). 
Ultimately, we average together only the measurements taken during the night when the temperature is more stable: a total of 750 integrations over 7 hours on the 8$^{th}$ and 9$^{th}$ of December, 2015.
Over this period the temperature changes by 0.25\,K.
Further details of this procedure are described in \citet{Monsalve2015}.

To smoothly interpolate the measurements over frequency, we use a \textsc{Polynomial} model, with a transformation of $\mathcal{T}(\nu) = \log(\nu/75\,{\rm MHz})$, parameters $p=1$ and $q=0$, and $n=10$ terms.
As with the other reflection coefficients, we first remove a delay before fitting the linear model, and restore the delay to the modeled solution.

\begin{figure}
    \centering
    \includegraphics[width=\linewidth]{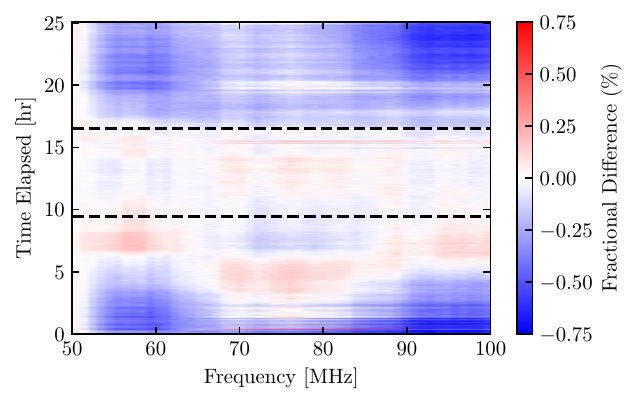}
    \caption{Stability of $\Gamma_{\rm ant}$ measurements over the observation time of 25 hours. Measurements used in the final averaged $\Gamma_{\rm ant}$ (night-time measurements) are between the black dashed lines. The heatmap represents the fractional difference in the \textit{real} part of the calibrated reflection coefficient, where the difference is taken with respect to the real part of the smoothed average (over the night-time observations), and the normalization is to the absolute value of the smoothed average. Stability is better than 0.5\% within the night-time measurements, with the bulk of observations staying within 0.2\% of the average.}
    \label{fig:ants11_stability}
\end{figure}

The stability of the $\Gamma_{\rm ant}$ measurements are illustrated in Fig. \ref{fig:ants11_stability}. 
In this figure, the quantity shown is the ratio of the difference between the real part of each calibrated measurement and the smoothed average to the absolute value of the smoothed average. 
Measurements remain within 0.2\% deviation for most of the observation window of 7 hours, with only a few outliers of $\sim 0.5\%$. 
While this measurement is also used to calibrate day-time data, after adjustment for the internal switch temperature the impact of using night-time vs. day-time measurements of $\Gamma_{\rm ant}$ is $<5$\,mK in the final calibrated average spectrum.

\subsubsection{Beam Chromaticity}
\label{sec:demo:workflow:beam}
To compute the beam factor chromaticity correction, we use the formalism described in \S\ref{sec:calibration:bfcc}, with the following choices.

For the sky model, we use the observations of \citet{Haslam1982}, represented on a regular grid of RA/dec, coupled with a spatially uniform spectral index of -2.5. 
For the beam model, we use a Feko simulation of the EDGES-2 low-band antenna described in \citet{Mahesh2021} (this beam simulation is also made available with the data provided at the link in the Data Availability section).
We use the alternative form of Eq. \ref{eq:beamfac}, i.e. with a frequency-dependent $T_{\rm sky}$, rather than defined at the reference frequency.
We use a reference frequency of 75\,MHz for the beam.

One subtlety is that the computation of the azimuth and elevation coordinates of the sky model pixels at a particular LST can be performed either with high-precision methods from \texttt{astropy} or instead with faster approximations based on rigid-body rotations (i.e. ignoring precession, nutation etc.). 
To make a close comparison with the legacy pipeline, we here use the faster approximation. However, this does not make a large impact on the spectral structure of the beam factor; its full effects will be explored in \mahesh.

Another small difference between this pipeline and the original is that instead of evaluating the `snapshot' beam factor on a different grid of LSTs for each observed day, we perform the calculation once on a fine LST grid covering the full 24 hours at a resolution of 6 minutes. 
We then interpolate the beam factor onto a daily-varying coarser grid of LSTs depending on the coverage of each day included with cubic splines. 
We discuss this further in \S\ref{sec:demo:workflow:datacal}.

\subsubsection{Data Calibration}
\label{sec:demo:workflow:datacal}
Having obtained a single averaged spectrum on each night of data as well as the quantities required to calibrate the data, we proceed now to apply those calibration solutions.
First, we down-select the spectrum channels to between $50-100$\,MHz (matching the calibration solutions). 
We also flag channels outside the range $51-99$\,MHz so that these do not contribute to any of the following model fits\footnote{The reason these are not just down-selected rather than flagged is that the calibration solutions are defined on the full $50-100$\,MHz range and so it is more convenient to apply them to the full range.}. We then simply $T_{\rm sca}$ and $T_{\rm off}$ from the calibration temperatures determined in \S\ref{sec:demo:workflow:rcvcal} and apply them to the spectra according to Eq. \ref{eq:TscaToff}\footnote{Note that to do this we first need to invert the calibration `guess' applied in \S\ref{sec:demo:workflow:singleday}.}. We also apply corrections for losses experienced through the balun and connector, $L_{\rm bnc}$, that connect the antenna to the receiver input, following \S\ref{sec:calibration:source-loss:antloss}.
In particular, the relevant properties of the balun and the connector are summarized in Table \ref{tab:bnc-properties}.
We set the ambient temperature $T_{\rm amb}=296$\,K for all observations, neglecting the ambient temperature recordings measured on site. 
The effect of using the recordings themselves will be investigated in \mahesh.

\begin{table}
    \centering
    \begin{tabular}{l|c|c}
         \textbf{Component} & \textbf{Balun} & \textbf{Connector} \\
         \hline
         Part Spec & - & SC3792 \\
         Outer Diameter [inch]  & 0.75 & 0.161 \\
         Inner Diameter [inch] & 5 / 16 & 0.05 \\
         Outer Material & Brass & Stainless Steel\\
         Inner Material & Copper & Copper\\
         Outer Conductivity [$5.96\times10^7$\,S/m] & 0.29 & 0.024 \\
         Inner Conductivity [$5.96\times10^7$\,S/m] & 1 & 0.24* \\
         Relative Permittivity of Dielectric & 1.07 & 2.05 \\
         Relative Interior Conductance & 0 \\
         Length [inch]  & 43.6 & 1.1811\\
         \hline
    \end{tabular}
    \caption{Properties of the balun and connector that determine their combined scattering parameters based on a physical model. While the inner material for the connector is copper, its conductivity is based on its plating, which is not copper, hence the lower conductivity of 24\% of the copper value.}
    \label{tab:bnc-properties}
\end{table}

We then apply the beam factor correction that was calculated in \S\ref{sec:demo:workflow:beam}. 
Here, we must average the beam factor over the LSTs observed in each particular day's observation.
Since the beam factor is relatively expensive to compute at a given LST, in B18 an approximation was made in which the beam factor was averaged over a regular grid of approximately 30 minute intervals, reducing the total required number of evaluations on any given observation day.
If the beam factor evolves slowly with LST, this should be reasonably accurate.
In B18, the precise extent of this grid was chosen such that the range of the grid was equal to the range of \textit{unflagged} data on any particular day, and the center of the grid was equal to the mean unflagged LST observed on that day. 
The grid resolution was set to be as close to 30 minutes as possible while dividing the extent evenly. 

In \ecode, we utilise an additional accelerator; we calculate the beam factor as a function of LST only once, on a finer grid over the full 24 hours of LST, and interpolate this result to the particular observations on each day. 
In principle, this enables us to interpolate directly to the observed, unflagged LSTs on each day with minimal computational overhead. 
Nevertheless, in this analysis, to remain as consistent as possible with B18, we interpolate from this fine resolution (6 minutes) grid to a per-day grid similar to B18 (i.e. $\sim30$,minute resolution). 
We note that there are some small differences between our 30-minute grid and the original from B18, whch arise due to the legacy code performing extra unnecessary conversions between time and LST, as well as using the timestamps of internal load+noise source spectra, rather than the antenna spectra.


To compute the final integrated beam factor, we use the method in which the numerator and denominator are averaged before taking their ratio.
Finally the beam factor is modeled with a 31-term Fourier series (c.f. Eq. \ref{eq:models:fourier}) with $\mathcal{T} = (\nu - 75\,{\rm MHz})/60\,{\rm MHz}$, and re-normalized to ensure that $G_{\rm beam}(75\,{\rm MHz}) = 1$.

In \S\ref{sec:demo:results} we show the effect of our slightly different approach to computing the beam factor on the ultimate averaged spectra, finding that interpolation from the universal 6-minute grid to an LST grid of 30\,min resolution approximating the legacy pipeline shows acceptable conformance to the original results.


In summary, the full calibration applied is:
\begin{equation}
    \hat{T}_{\rm obs} = G^{-1}_{\rm beam}(1-L_{\rm bnc})^{-1} \left(\hat{T}_{\rm sca} Q + \hat{T}_{\rm off} - T_{\rm amb}L_{\rm bnc}\right),
\end{equation}
with $T_{\rm amb} = 296$\,K.

After applying this calibration, we further down-sample the frequency axis.
We again use a Gaussian kernel, and down-sample by a factor of eight, resulting in a frequency resolution of $\approx 390$\,kHz.
However, in this case a more careful treatment of non-uniform weights is required.
We use Eq. \ref{eq:unbiased-average}, where the model used is \textsc{LinLog} with $n=3$ terms, $\beta=-2.5$ and $\mathcal{T}(\nu) = \nu/75\,$MHz.
The weights used in the fit are uniform except that flagged channels are respected.
The residuals to the model are convolved with the kernel, and channels surrounded by at least four flagged channels to either side are re-flagged.

\subsubsection{Average Over days}
\label{sec:demo:workflow:nightavg}
The last major analysis step brings the spectra from each night together, performing a few final quality checks before averaging over all days.
Before averaging, we require a new model of the data so that Eq. \ref{eq:unbiased-average} can be applied correctly.
Here, we fit a 5-term \textsc{LinPhys} model, with $\beta=-2.55$ and $\mathcal{T}(\nu) = \nu/75\,$MHz.
We use the full accumulated per-channel number of samples of each night's spectrum as the weights in this fit.

Our first filter is to remove all days for which the weighted RMS of residuals to this physical model over the full frequency range is greater than 170\,mK.
This is an extremely conservative filter, and may be problematic in some ways.
In particular, if the foreground and signal models were uncorrelated (which is not true), this filter would flag all days if the signal itself had an amplitude greater than 170mK.
Furthermore, setting the threshold to 170\,mK is clearly dependent on the noise-level we expect from each day, given the number of integrations included. 
We will explore the ramifications of this filter further in \mahesh.
Ultimately, this conservative filter removes 76 of the 138 days (more than 50\%).

\begin{figure}
    \centering
    \includegraphics[width=\linewidth]{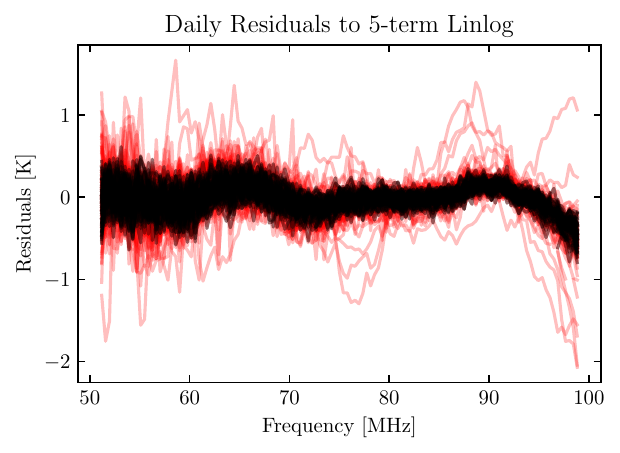}
    \caption{Residuals of each day's averaged and calibrated spectrum to a 5-term \textsc{LinLog} model. Black curves indicate the days retained after the RMS filter (c.f. \S\ref{sec:demo:workflow:nightavg}), while red are their complement.
    The general shape of the final foreground-residuals of the fully-averaged data is already clearly visible on each day. The retained days are consistent to well within the noise level. 
    }
    \label{fig:daily-residuals}
\end{figure}

Figure Fig. \ref{fig:daily-residuals} illustrates the consistency of the spectra from each day at this stage of analysis. In the plot the black curves are the days retained after the RMS filter, while the red are those thrown out. 
Already the general form of the final foreground residual to the fully-averaged data is apparent on each day, and the retained days are consistent to well within the noise level. 

\begin{figure}
    \centering
    \includegraphics[width=\linewidth]{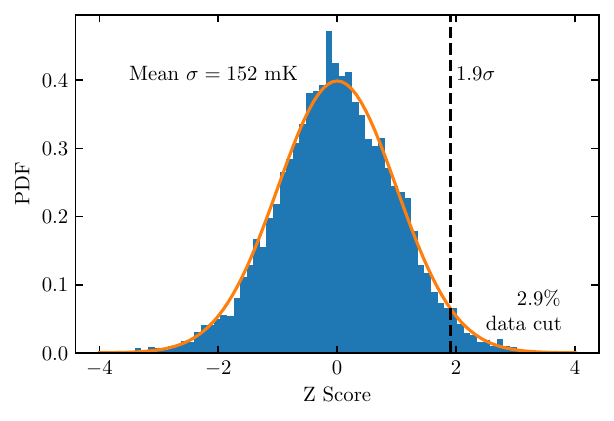}
    \caption{Histogram of $Z$-scores for all channels and days after removing inconsistent days (i.e. red data in Fig. \ref{fig:daily-residuals}). $Z$-scores are defined as residuals to a \textsc{LinPhys} model divided by the RMS of these residuals over the full spectrum. The average RMS over all days is 152\,mK. The vertical dashed line marks the threshold for the second round of RFI flagging.}
    \label{fig:noise-distribution}
\end{figure}

The second filter is a second deeper check for RFI.
Here, we flag any channel whose residual to the physical foreground model is greater than $1.9 \sigma$, where $\sigma$ is the RMS of residuals to the \textsc{LinPhys} model of the full spectrum for that night\footnote{This can be cast in the framework of the general iterative RFI filter as using \textsc{LinPhys} model to generate the data model, and a top-hat kernel with size twice that of the spectrum to model the standard deviation, along with a maximum of one iteration, and no watershed flagging.}.
Again, a threshold of 1.9 is conservative, and may distort the gaussianity of the thermal noise in the final averaged spectrum by clipping truly in-distribution channels. 
Indeed, Fig. \ref{fig:noise-distribution} indicates that the data are well-characterized by a normal distribution with a constant standard deviation given by the RMS over the full spectrum, and that this cut is merely excising in-distribution noise fluctuations.
This filter flags about 3\% of the remaining data.

Following these quality checks, we average the spectra from all days together, using Eq. \ref{eq:unbiased-average}. 
Here, note that we do \textit{not} re-fit a model spectrum to each night (using the new flags), however we do use these updated flags to weight the average. 

\begin{figure}
    \centering
    \includegraphics[width=\linewidth]{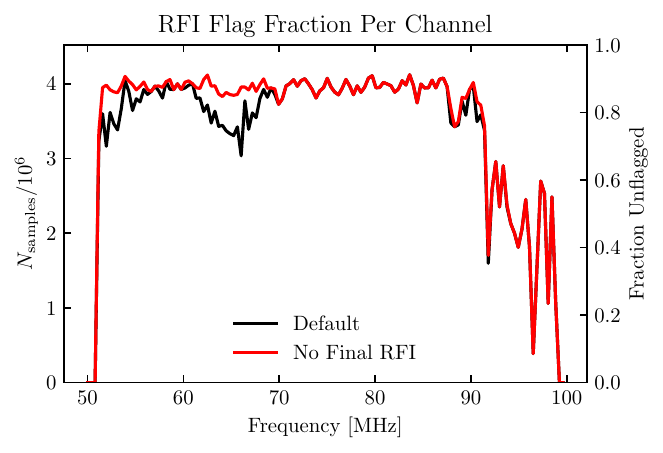}
    \caption{Number of samples in each fully-averaged channel, where each sample corresponds to one raw integration ($t_{\rm intg} \approx 13\,sec$) at raw frequency resolution $\Delta \nu = 6.1\,{\rm kHz}$). The right-hand axis scales these to the total number of possible samples (discounting fully-flagged integrations) to yield an `unflagged fraction'. Note that due to the gaussian kernel applied when down-sampling channels, neighbouring channels are not quite independent (so the number of samples is a slight over-estimate for estimating joint data uncertainty).}
    \label{fig:nsamples}
\end{figure}

The number of samples in each fully-averaged channel is shown in Fig. \ref{fig:nsamples}.
Here, each `sample' corresponds to one raw integration ($t_{\rm intg} \approx 13\,{\rm s}$) at raw frequency resolution $\Delta \nu = 6.1\,{\rm kHz}$). The right-hand axis of Fig. \ref{fig:nsamples} scales $N_{\rm samples}$ to the total number of possible samples (discounting fully-flagged integrations) to yield an `unflagged fraction'. 
There are two obvious regions that are preferentially flagged: the FM band above 90\,MHz and a region from 60-70\,MHz.
This latter region does not correspond to RFI flagged in the initial per-night quality checks (c.f. \S\ref{sec:demo:workflow:singleday}), as demonstrated by Fig. \ref{fig:rfi-occupancy}.
The only other per-channel flagging performed is the second RFI flagging step that occur after calibration (c.f. \S\ref{sec:demo:workflow:nightavg}), which uses a much less sophisticated algorithm (no iteration, and lower-order data model), with a much stricter threshold of 1.9\,$\sigma$.
Given that the data model used for this flagging does not include a 21\,cm component, it might be expected that the model would be systematically biased by more than 1.9\,$\sigma$ at certain frequency ranges, resulting in potentially spurious flags. 
Indeed, removing this final RFI flagging step, we obtain the red line in Fig. \ref{fig:nsamples}, for which the sampling is reasonably uniform at $\sim 90\%$ except for within the FM band.

\subsection{Results}
\label{sec:demo:results}

Within the pipeline discussed throughout this section, two components were identified for which \ecode\ intentionally does not match the legacy pipeline. 
In both cases the choice was made because the legacy pipeline had a less robust solution, and implementing that solution within the new pipeline (as a non-default option) was either impossible or so intricate as to make the effort not worthwhile.
These two components are the first round of RFI flagging on a nightly basis (\S\ref{sec:demo:workflow:singleday}) and the computation of the integrated beam factor (\S\ref{sec:demo:workflow:datacal}). 

\begin{figure}
    \centering
    \includegraphics[width=\linewidth]{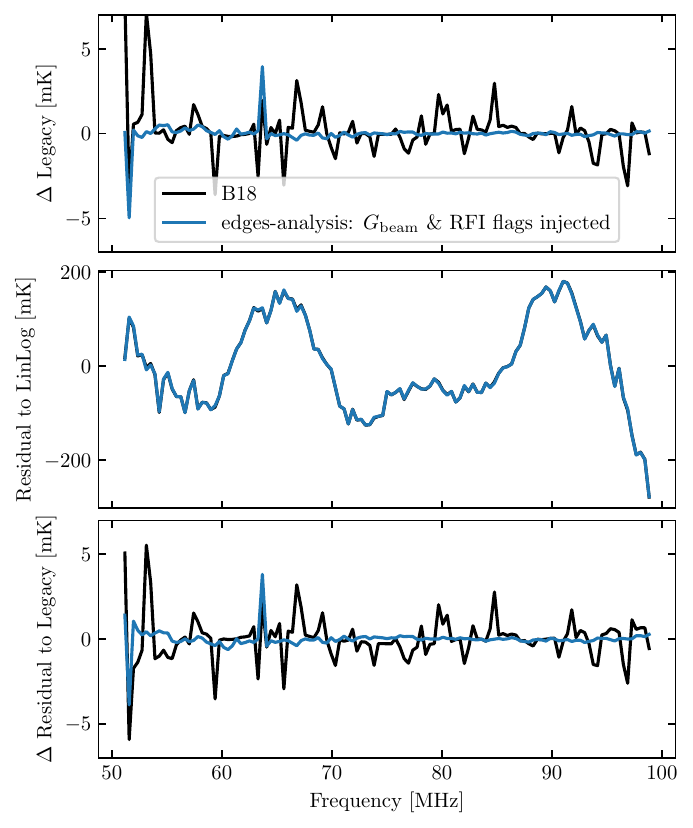}
    \caption{Comparison of averaged spectrum to legacy code. Blue lines show spectra produced with \ecode\, and black lines the published spectrum of B18 (which uses the legacy code with one small bug, see text). The top panel shows the absolute difference between the spectra, the middle panel the residual to a 5-term \textsc{LinLog} foreground model, and the lower panel the difference between the foreground-residuals (e.g. residuals of \ecode\ minus residuals of legacy). These results indicate that the numerical differences between \ecode\ and the legacy pipeline (neglecting $G_{\rm beam}$ and RFI flags) are smaller than the differences produced by the small fixed bug.}
    \label{fig:injected-result-compare}
\end{figure}

In both cases, in order to check that the rest of the pipeline is accurate, a temporary measure is to \textit{inject} the output of the legacy pipeline for these particular components (the per-day RFI flags and the per-day beam factor), and run the rest of the pipeline as normal. 
The result of this exercise is shown in Fig. \ref{fig:injected-result-compare}.
Note that here and elsewhere in this section, we compare to the final averaged spectrum of the `legacy' pipeline, which we have re-processed for the purpose of this work.
This is not \textit{exactly} the same as the H2 case from \bowman\ (which was made publicly available at the time) because a single very minor bug was found in that code which has since been fixed.
This bug was simply that the first channel was being ignored when fitting models to the reflection coefficients (\S\ref{sec:demo:workflow:rcvcal}). 
For the sake of completeness, we also compare the original public spectrum to that re-processed with the bug-fixed legacy pipeline in  Fig. \ref{fig:injected-result-compare} (black lines).
When injecting $G_{\rm beam}$ and RFI flags, \ecode\ produces a final spectrum in excellent agreement with legacy pipeline---in fact, it is in better agreement than the public results from B18, at much better than 1\,mK over the band.

\begin{figure}
    \centering
    \includegraphics[width=\linewidth]{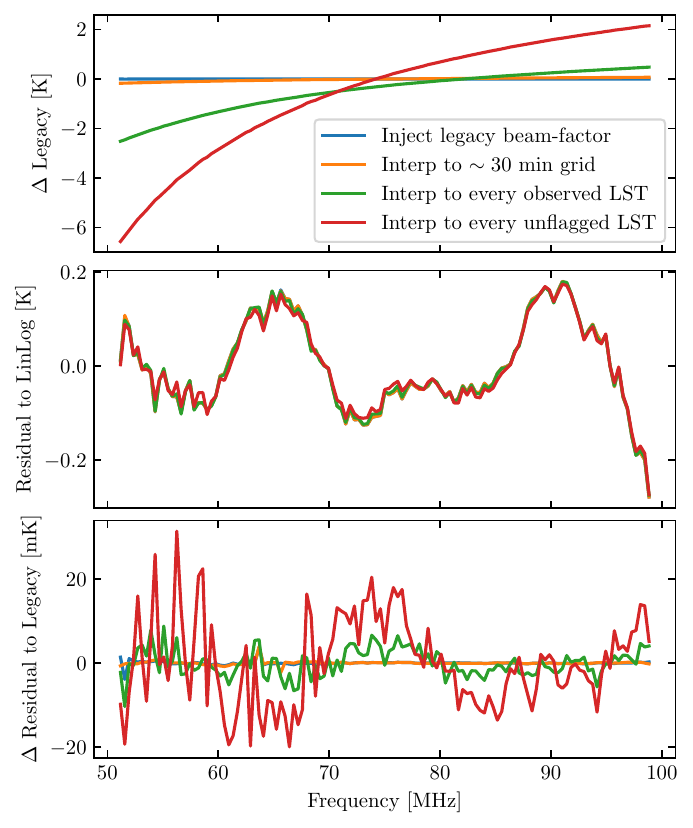}
    \caption{Comparison of the effect of beam-factor choices on the final averaged spectrum. Panels are the same as Fig. \ref{fig:injected-result-compare}, but each colour (other than blue) represents a different method of assigning the LSTs over which the beam factor is averaged.}
    \label{fig:final-comparison-beam-factor}
\end{figure}

Having established an accurate reproduction with the injection of the two analysis components that were intentionally modified, we now examine the impact of using the more robust implementation of each of these components within \ecode.
Fig. \ref{fig:final-comparison-beam-factor} compares the new implementation of the beam factor, using the same quantities as Fig. \ref{fig:injected-result-compare}. 
Recall that the primary difference in the new implementation is that we evaluate the beam factor on a fine LST grid and interpolate it to the LSTs represented on each day(c.f \S\ref{sec:demo:workflow:beam}).
In blue we show the averaged spectrum with the beam factor from the legacy pipeline injected (i.e. the same as the blue line in Fig. \ref{fig:injected-result-compare}). 
In orange, we show our fiducial case, in which we interpolate from our fine LST grid to a $\sim30$-minute grid similar to that used in B18.
Here the residuals are smooth and at a level of a couple of mK.
Green and red show cases that may be better motivated than the legacy pipeline, in which the LSTs to which the beam factor is interpolated on each day match either all the observed LSTs (green) or all \textit{unflagged} LSTs (red) on any given day. These two result in one or more full days being flagged differently later in the pipeline, resulting in an absolute difference of a few K (top panel), but the differences are mostly absorbed by the foregrounds, such that the difference between foreground residuals is below 20 mK (bottom panel).

\begin{figure}
    \centering
    \includegraphics[width=\linewidth]{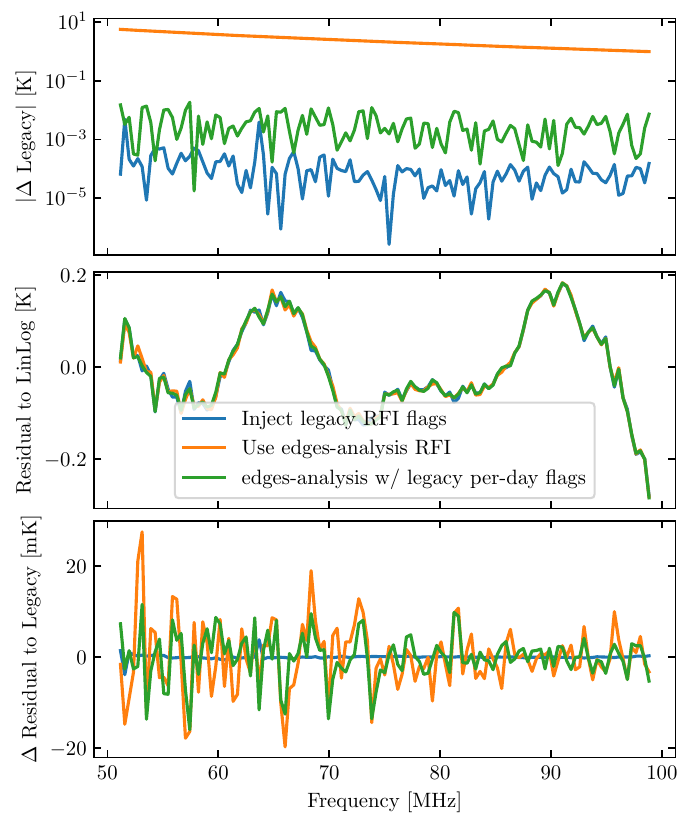}
    \caption{Comparison of the effect of RFI algorithm on the final averaged spectrum. Panels are the same as Fig. \ref{fig:injected-result-compare}. Blue line uses injected flags from the legacy pipeline (identical to Fig. \ref{fig:injected-result-compare}). Orange line uses \ecode\ with settings described in \S\ref{sec:demo:workflow:singleday}. Green line also uses \ecode\ but ensures that the days used in the final spectrum match the legacy pipeline.}
    \label{fig:xrfi_result_compare}
\end{figure}

Now we consider the effect of using \ecode\ to determine nightly RFI flags. 
Fig. \ref{fig:xrfi_result_compare} compares three cases to the legacy pipeline: (i) injecting legacy RFI flags directly (i.e. the same as in Fig. \ref{fig:injected-result-compare}), (ii) using \ecode\ to identify flags, as per \S\ref{sec:demo:workflow:singleday}, and (iii) using \ecode\  again, but ensuring that the final list of days included in the average matches the legacy pipeline. 
We find that using \ecode\ results in a large ($>$4\,K) absolute discrepancy, though after fitting and subtracting a foreground model the discrepancy is very small ($< 10$\,mK) and noise-like.
The absolute discrepancy is almost entirely due to the fact that finding different per-channel flags on each night results in a few days being omitted/included differently than the legacy pipeline due to the RMS threshold (c.f. \S\ref{sec:demo:workflow:nightavg}). 
Accounting for this effect using case~(iii; green line) brings the absolute difference down to $\lesssim 10$\,mK.
Notably, there are no large spikes in the bottom panel of Fig. \ref{fig:xrfi_result_compare}, indicating that the channels that are flagged differently do not have strong or consistent RFI (the differences are noise-like). 

\begin{figure}
    \centering
    \includegraphics[width=\linewidth]{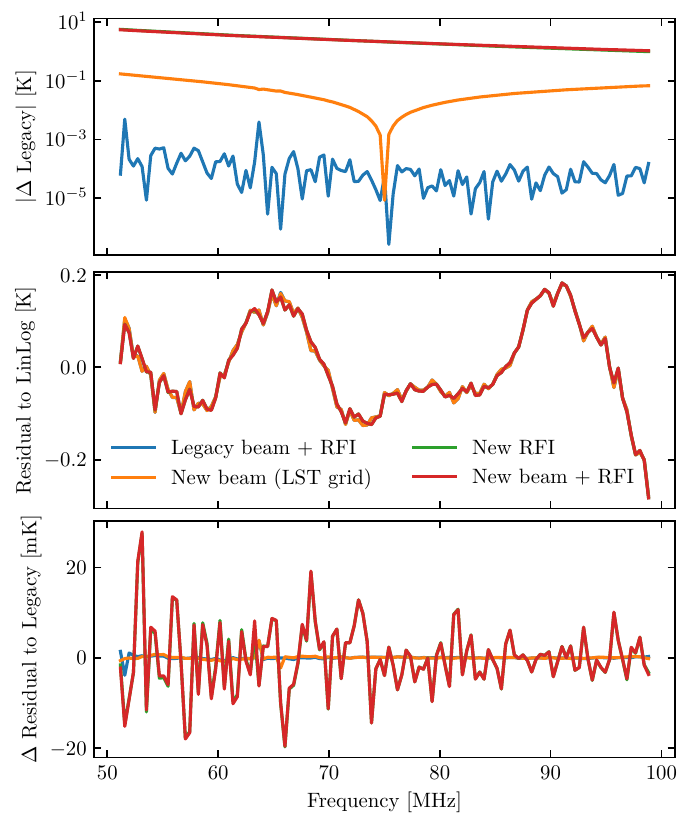}
    \caption{Comparison of the final averaged spectrum between the new \ecode\ pipeline and the legacy pipeline. Panels are the same as Fig. \ref{fig:injected-result-compare}. We show the progression from injecting both the legacy beam and RFI flags, to performing each of these separately with \ecode\ (c.f Figs. \ref{fig:final-comparison-beam-factor} and \ref{fig:xrfi_result_compare}), to our final result in which both of these components are computed with the new pipeline (red lines).}
    \label{fig:final-comparison}
\end{figure}

Fig. \ref{fig:final-comparison} shows the results of running the full pipeline detailed in the previous section, compared to the output of the legacy pipeline.
In this figure, we show the full progression of changes we have just outlined: (i, blue) injection of both beam factor and RFI flags from the legacy pipeline, (ii, orange) using \ecode\ to compute the beam factor (on the 30\,min grid approximating the legacy pipeline, i.e. the green line from Fig. \ref{fig:final-comparison-beam-factor}), (iii, green) using \ecode\ to identify RFI flags (i.e. the orange line from Fig. \ref{fig:xrfi_result_compare}) and finally (iv, red) using \ecode\ for both components, i.e. our fiducial reproduction of the results of B18 using the new pipeline.
The largest difference is made by the identification of RFI, but as already discussed, this effect is dominantly a smooth component that is fit out by the foreground model plus a noise-like component.

\subsection{Discussion and Future Considerations}
\label{sec:demo:discussion}
Throughout the preceding demonstration, in which we explicitly reproduced the results of \bowman, several specific aspects of the analysis were identified as likely suboptimal. Here we summarize these aspects to facilitate future improvements.

The only thing considered a true \textit{bug} in the legacy pipeline was the erroneous omission of a single channel (the lowest channel) when fitting reflection coefficients (noted in \S\ref{sec:demo:results}). 
This bug has a very minor effect (c.f. Fig. \ref{fig:injected-result-compare}) and has been fixed in the legacy code (all results in this work compare to the fixed version). Some other accuracy-related points noted were (i) the usage of approximate astrometry, ignoring second-order effects such as precession and nutation of Earth's rotation; (ii) use of an older version of the Haslam sky model \citep{Haslam1982} in regular RA/Dec coordinates for which the solid angle element in the sky integral is only computed approximately; (iii) use of nearest-grid-point interpolation of the beam when evaluating the beam chromaticity; (iv) use of a simplistic foreground model with a spatially-invariant spectral index when computing the beam chromaticity.
It is not expected that these approaches have a marked effect on the spectral structure of the averaged spectrum, but it is easy to use more accurate versions of all of these in \ecode.

A more serious point---which nevertheless seems not to have a strong impact---is that the legacy pipeline does not reduce the condition number of linear systems when performing linear modeling (c.f. \S\ref{sec:demo:workflow:singleday}), which affects the modeling of spectra and reflection coefficients and also the determination of RFI flags. 
We have demonstrated that linear equation solving in \ecode\ is more robust, and should be used in future applications.

The remaining considerations have to do with analysis choices, especially around data quality checks:
\begin{itemize}
    \item It is assumed that each calibration source is at an ambient temperature of 296\,K except for the hot load, which is at 399\,K. Since measurements of the ambient temperature are taken while running the calibration spectra, it would be better to use this information directly.
    \item In a similar vein, the analysis uses an ambient temperature of 296\,K to compute the antenna loss, instead of utilizing the measurements from the field where they are available.
    \item In the filter for strong RFI spikes in the FM range (filter (iii) in \S\ref{sec:flagging:integrations:spectrum}), the threshold must be somewhat arbitrary, and its effect will change in different conditions (e.g. when the galaxy is overhead the temperature and noise level are higher). Since the purpose of this filter is to guard against non-linearities, it would be more appropriate for the metric to be more closely tied to the conditions that cause non-linearities.
    \item The filter for fractional power in a specific frequency range (filter (v) in \S\ref{sec:flagging:integrations:spectrum}) must be carefully tuned to the data and telescope under consideration. In its current form, demanding that the power in the upper half of the band (100-200\,MHz) be between 0.7 and 3\% of the total power is not transparently based on any physical requirements, and this should be replaced by a filter that targets specific unique failure modes. 
    \item We noted a couple of potential issues with the per-night filter which throws out days whose RMS residuals to a \textsc{LinPhys} is greater than 170\,mK (c.f. \S\ref{sec:demo:workflow:nightavg}), including that it could throw out all days due to a cosmological signal of requisite amplitude. An alternative method to filter out discrepant days before final averaging would be to consider \textit{relative} discrepancy with other days.
    \item Fig. \ref{fig:nsamples} illustrates that the second round of RFI flagging may be too aggressive (with a 1.9\,$\sigma$ threshold), and end up being dominated by the insufficiency of the foreground model rather than true RFI. One way to improve this flagging would be to use a more flexible data model or a more careful iterative algorithm (c.f. \S\ref{sec:flagging:rfi:general-iterative}).
    \item Following the second round of RFI flagging (which occurs on the calibrated averaged spectrum for each night), it would make sense to re-fit a data model before averaging each night together, in order to have the least biased average possible (c.f. Eq. \ref{eq:unbiased-average}).
\end{itemize}

Many of these shortcomings will be examined in \mahesh.

\section{Conclusions} \label{sec:conclusions}
We have for the first time presented a comprehensive set of analysis methods used in the EDGES-2 global 21\,cm experiment, including quality assurance and data flagging, calibration, modeling and averaging. 
At the same time, we have presented a new suite of open-source software that implements these methods in a clear and reproducible way.
As an application of these methods, we have given an explicit account of the analysis that produced the result of \bowman, and have made the data and pipeline required to reproduce this result public. This software consists of a core data interface, \gscode, which is a powerful and reliable interface to arbitrary single-antenna 21\,cm data, as well as an analysis package, \ecode, that provides a host of well-tested analysis methods that might be cross-applied to datasets from different experiments.

The detailed presentation of the specific analysis pipeline leading to the results of \bowman\ (\S\ref{sec:demo}) uncovered several choices that were not comprehensively justified, including the LSTs over which the beam chromaticity factor is integrated, various thresholds used for flagging integrations and channels, and the precise set of VNA measurements and calibrations used for lab-based calibration.
Our software framework makes it easy to investigate the effects of modifying these choices, and in \mahesh\ we will present just such an exploration.
Nevertheless, we showed that the new pipeline can be used to reproduce the results of B18 to within a few mK, and that the differences are well understood. 

We hope that this work helps the community both to understand and examine this important result, as well as to converge on a common analysis framework across experiments, in order to accelerate progress through easier sharing of data and methodologies.

\section*{Data Availability}
The data required to reproduce the \textit{calibration} of the EDGES-2 receiver used here and in \bowman\ is located at \url{https://doi.org/10.5281/zenodo.18091240}.
The raw field spectra used in \bowman\ and in this work are located at \url{https://loco.lab.asu.edu/edges/edges-data-release/}.
The precise pipeline used to reproduce the results of \bowman\ with the software described here can be found at \url{https://doi.org/10.5281/zenodo.20215409} (which also contains evaluated Jupyter notebooks of all pipeline stages).

\begin{acknowledgments}
SGM has received funding from the European Union’s Horizon 2020 research and innovation programme under the Marie Skłodowska-Curie grant agreement No 101067043.
This research was funded by the following National Science Foundation grants: AST-1609450, AST-1813850, and AST-1908933.  
EDGES is located at the Inyarrimanha Ilgari Bundara, the CSIRO Murchison Radio-astronomy Observatory.  We acknowledge the Wajarri Yamatji people as the traditional owners of the Observatory site. 
We also thank CSIRO for providing site infrastructure and on-going support. 
\end{acknowledgments}

\facility{EDGES}


\software{
\texttt{numpy} \citep{Harris2020},
\texttt{scipy} \citep{Virtanen2020},
\texttt{matplotlib} \citep{Hunter2007},
\texttt{astropy} \citep{AstropyCollaboration2013,AstropyCollaboration2018}, \texttt{jupyter} \citep{Kluyver2016jupyter},
\texttt{HDF5} \citep{the_hdf_group_2025_17808614}
\texttt{nextflow} \citep{DiTommaso2017}
}





\appendix

\section{Cascading Two-Port Networks}
\label{sec:cascading}
If two or more two-port networks are joined together (`cascaded'), the resulting $S$-parameters of the full network can be described by first transforming to the $T$-parameters (`transfer parameters') \citep{dunsmore2020handbook}
\begin{equation}
    \mathbf{T} = \begin{pmatrix}
        -|\mathbf{S}|/\mathbf{S}_{21} & \mathbf{S}_{11}/ \mathbf{S}_{21} \\
        -\mathbf{S}_{22}/\mathbf{S}_{21} & \mathbf{S}_{21}^{-1},
    \end{pmatrix}
\end{equation}
then computing the matrix product of all $\mathbf{T}$:
\begin{equation}
    \mathbf{T}_{\rm cascaded} = \mathbf{T}_1 \mathbf{T}_2\mathbf{T}_3 \cdots,
\end{equation}
and finally converting back to an $\mathbf{S}$-matrix:
\begin{equation}
    \mathbf{S} = \begin{pmatrix}
        \mathbf{T}_{12}/\mathbf{T}_{22} & |\mathbf{T}|/\mathbf{T}_{22} \\
        \mathbf{T}_{22}^{-1} & -\mathbf{T}_{21}/\mathbf{T}_{22}
    \end{pmatrix}.
\end{equation}

\section{Cross-reference of Calibration Quantities with REACH Bayesian Pipeline}
\label{app:calibration-comparison}

Note that the calibration quantities as encoded in this paper may be related directly to those expressed in other works.
In particular, the linear coefficients $X_i$ defined in \citet{Roque2021} (R21; see also \citep{Kirkham2025,Kirkham2026,Dasgupta2026b}) are related to our quantities as follows:
\setlength{\extrarowheight}{10pt}
\begin{equation*}
\begin{array}{c >{\displaystyle}c>{\displaystyle}c}
     {\bf Notation (R21)} & {\bf Formula (Here)} & {\bf Formula (R21)}  \\
     \hline 
     X_{\rm L} & \epsilon_{\rm src}^{-1}(1 - |\Gamma_{\rm rcv}|^2) & \frac{|1-\Gamma_{\rm rcv}\Gamma_{\rm src}|^2}{1 - |\Gamma_{\rm src}|^2}   \\
    X_{\rm unc} & -\epsilon^{-1}_{\rm src}K_0 &   -\frac{|\Gamma_{\rm src}|^2}{1 - |\Gamma_{\rm src}|^2} \\
    X_{\rm cos} & -\epsilon^{-1}_{\rm src}K_1&  -\mathrm{Re}\left(\frac{\Gamma_{\rm src}}{1 - \Gamma_{\rm src}\Gamma_{\rm rcv}} \times \frac{X_L}{1 - |\Gamma_{\rm rcv}|^2}\right) \\
    X_{\rm sin} & -\epsilon^{-1}_{\rm src} K_2 &  -\mathrm{Im}\left(\frac{\Gamma_{\rm src}}{1 - \Gamma_{\rm src}\Gamma_{\rm rcv}} \times \frac{X_L}{1 - |\Gamma_{\rm rcv}|^2}\right) \\
    X_{\rm NS} & \epsilon^{-1}_{\rm src} Q_{\rm src}(1 - |\Gamma_{\rm rcv}|^2) & Q_{\rm src} X_L \\
    \hline
\end{array}    
\end{equation*}
\setlength{\extrarowheight}{0pt}

\section{Concise Summary of Pipeline}
\label{app:pipeline-table}

In Table \ref{tab:b18-pipeline} we present a concise overview of the pipeline steps involved in reproducing the results of \bowman, as discussed in \S\ref{sec:demo}.

\begin{deluxetable}{lcll}
\tablewidth{\linewidth}
\tablecolumns{4}
\tablehead{
    \colhead{Name} & \colhead{Type} & \colhead{Parameters} & \colhead{Notes}
}
\tablecaption{Concise table of pipeline steps for the demonstration of \S\ref{sec:demo}.\label{tab:b18-pipeline}}
\startdata
     \sidehead{\textbf{Per-Night Filtering and Averaging. \S\ref{sec:demo:workflow:singleday}. Input: Raw Spectra from 24 hours of observation}.}
     Galactic Centre & Filter & Keep LST=23.76-11.76 hr & \S\ref{sec:flagging:integrations:auxiliary}; \S\ref{sec:demo:workflow:singleday}\\
     Auxiliary Data & Filter & ADC $< 0.35\,{\rm V}^2{\rm Hz}^{-1}$& \S\ref{sec:flagging:integrations:auxiliary}; \S\ref{sec:demo:workflow:singleday}. ``ADC''. \\
     Power Percent & Filter & $0.7 \% < {\rm pp} < 3\%$ & \S\ref{sec:flagging:integrations:spectrum} (v); \S\ref{sec:demo:workflow:singleday}, ``PPF''. \\
     Dicke Calibration & Calibration & & Eq. \ref{eq:dickecal} \\
     Scale to Approx. Temp. & Calibration & $T_{\rm sca}=1000{\rm K}$; $T_{\rm off}=300\,{\rm K}$ & Eq. \ref{eq:TscaToff} \\
     Peak Power & Filter & $P_{137-138\,{\rm MHz}}/P_{80-200\,{\rm MHz}} < 40$ & \S\ref{sec:flagging:integrations:spectrum} (ii); \S\ref{sec:demo:workflow:singleday}. ``OrbComm''. No flags.\\
     Single Channel Spike & Filter & 88-120\,MHz; 200\,K & \S\ref{sec:flagging:integrations:spectrum} (iii); \S\ref{sec:demo:workflow:singleday}. ``FM''\\
     RMS & Filter & 60-80\,MHz; $m \propto \nu^{-2.5}$; Threshold 200\,K. & \S\ref{sec:flagging:integrations:spectrum} (iv); \S\ref{sec:demo:workflow:singleday}. ``RMS''. \\
     Average Integrations & Average & Uniform Weights; All Unflagged times per day & Eq. \ref{eq:unbiased-average}. \\
     Select Freqs & Filter & $40-100$\,MHz & \\
     RFI Excision I & Filter & Model: 37-term \textsc{Fourier}; &  \S\ref{sec:flagging:rfi:sliding} \\
     & & STD: 614-channel box-car; & \\
     & & Threshold 2.5$\sigma$ & 
     \\
     Frequency Down-Sample & Average & 8-channel; Gaussian Kernel & \S\ref{sec:averaging} \\
     \sidehead{\textbf{Per-Night Calibration. \S\ref{sec:demo:workflow:datacal}. Input: single uncalibrated averaged spec. from one day.}}
     Select Freqs & Filter & 51-99 MHz &  \\
     Receiver Calibration & Calibration & See \S\ref{sec:demo:workflow:rcvcal} & \S\ref{sec:calibration}\\
     Balun+Connector Correction & Calibration  & Table \ref{tab:bnc-properties} & \S\ref{sec:calibration:source-loss:antloss}\\
     Beam Chromaticity Correction & Calibration & See \S\ref{sec:demo:workflow:beam} & \S\ref{sec:calibration:bfcc} \\
     Frequency Down-Sample & Average & 8-channel; Gaussian Kernel & \S\ref{sec:averaging} \\
     \sidehead{\textbf{Night-to-night averaging. \S\ref{sec:demo:workflow:nightavg}. Input: Single calibrated averaged spec. from all days.}}
     Per-night RMS & Filter & 5-term \textsc{LinPhys}; $\beta=-2.55$; Thresh. 170\,mK & Eq. \ref{eq:models:linphys}; \S\ref{sec:demo:workflow:nightavg} \\
     RFI Excision II & Filter & 5-term \textsc{LinPhys}; $\beta=-2.55$; Thresh. 1.9$\sigma$ & \ref{eq:models:linphys}; \S\ref{sec:flagging:rfi}; \S\ref{sec:demo:workflow:nightavg}\\
     Average over nights & Average & Flagged-Uniform Weighting & Eq. \ref{eq:unbiased-average}; \S\ref{sec:demo:workflow:nightavg} \\
\enddata
\tablecomments{These are the pipeline steps for the analysis that reproduces \citet{Bowman2018}, as discussed in \S\ref{sec:demo}. Only steps applied to field data are shown, see Fig. \ref{fig:b18-pipeline} for a schematic showing ancillary steps (such as computing the receiver calibration). Major groupings of steps marked by bold headers indicate different Jupyter notebooks in the implemented pipeline. Steps are applied in order from top to bottom.}

\end{deluxetable}

\bibliographystyle{aasjournalv7}
\bibliography{library} 



\end{document}